\documentclass[useAMS,usenatbib,usegraphicx]{mn2e}
\usepackage{color}
\usepackage{aas_macros}
\usepackage{amsmath}
\usepackage{amssymb}
\usepackage{graphicx}
 \usepackage{multirow}
 \usepackage{float}
\usepackage{subfig}
\title[]{SILCC-ZOOM: The early impact of ionizing radiation on forming molecular clouds}
\author[S. Haid]{S. Haid$^{1}$\thanks{E-mail:
haid@ph1.uni-koeln.de}, S. Walch$^{1}$, D. Seifried$^{1}$, R. W\"unsch$^{2}$, F. Dinnbier$^{1}$ and T. Naab$^{3}$\\
$^{1}$I. Physikalisches Institut, Universit\"at zu K\"oln, Z\"ulpicher-Strasse 77, 50937 Cologne, Germany\\
$^{2}$Astronomick\'{y} \'{U}stav, Akademie v\u{e}d \u{c}esky Republiky, Bo\u{c}n\'{i} II 1401, CZ-14131 Praha, Czech Republic\\
$^{3}$Max-Planck-Insitut f\"ur Astrophysik, Karl-Schwarzschild-Strasse 1, 85741 Garching, Germany}
\begin{document}

\maketitle

\label{firstpage}

\begin{abstract}
As part of the SILCC-ZOOM project we present our first sub-parsec resolution radiation-hydrodynamic simulations of two molecular clouds self-consistently forming from a turbulent, multi-phase ISM. The clouds have similar initial masses of few $10^4\;{\rm M}_{\odot}$, escape velocities of $\sim$5 km s$^{-1}$, and a similar initial energy budget. We follow the formation of star clusters with a sink based model and the impact of radiation from individual massive stars with the tree-based radiation transfer module \textsc{TreeRay}. Photo-ionizing radiation is coupled to a chemical network to follow gas heating, cooling and molecule formation and dissociation. For the first 3 Myr of cloud evolution we find that the overall star formation efficiency is considerably reduced by a factor of  $\sim$4 to global cloud values of $<$ 10 \% as the mass accretion of sinks that host massive stars is terminated after $\lesssim$1 Myr. Despite the low efficiency, star formation is triggered across the clouds. Therefore, a much larger region of the cloud is affected by radiation and the clouds begin to disperse. The time scale on which the clouds are dispersed sensitively depends on the cloud substructure and in particular on the amount of gas at high visual extinction. The damage of radiation done to the highly shielded cloud (MC$_{1}$) is delayed. We also show that the radiation input can sustain the thermal and kinetic energy of the clouds at a constant level. Our results strongly support the importance of ionizing radiation from massive stars for explaining the low observed star formation efficiency of molecular clouds.

\end{abstract}


\section{Introduction}
\label{sec-introduction}

Molecular clouds (MC) condense out of the diffuse, interstellar medium (ISM). These dense regions host filamentary substructures of molecular gas with an atomic envelope \citep{andre14, dobbs14, klessen16}. Massive stars form in infrared dark clouds, which are the densest parts of MCs \citep{goldreich74, lada03, Rathborne2006, klessen11, Ragan2012}. During their lifetime, massive stars emit ionizing radiation and eject high-velocity winds, which result in the deposition of momentum, kinetic, and thermal energy in the ISM and change the chemical composition. The underlying physical processes are collectively termed \textit{stellar feedback}, i.e. stellar winds \citep{castor75, weaver77, wunsch11}, ionizing radiation \citep{spitzer78, hosokawa06, dale12, walch12}, radiation pressure \citep{krumholz09b, fall10, murray10}, and supernovae (SNe, \citealt{sedov58, ostriker88, walch15a, koertgen16}). Feedback modifies density structures, counteracts the gravitational collapse, interrupts mass accretion, and directly influences the cycle of star formation. However, the detailed impact on molecular cloud evolution is still a matter of discussion \citep{whitworth79, krumholz06, krumholz09c, walch12, dale15}. It seems clear that stellar feedback can change the local and global multi-phase structure of the ISM with dramatic consequences for star formation \citep{naab17}.

MCs are complex.  Observations indicate that they are embedded in their galactic environment \citep{maclow04} and coupled to large scale (some 100 pc) motions \citep{hughes13, colombo14}. Galactic turbulent velocity fields seem to be inherited \citep{brunt09} with consequences for the star formation rate \citep{reyraposo15}. Hence, it is likely that the cloud properties are already imprinted during early formation and continuously reshaped by physical processes on galactic scales \citep{dobbs12, walch15, girichidis16, seifried17, reyraposo17}. This also suggests that the possible support of MCs by internal stellar feedback is highly variable and depends on the cloud structure. Analytical models which are usefully guiding our theoretical understanding, may not fully reflect the complexity of self-consistently evolving MCs \citep{matzner02}. 

Early studies treat MCs in isolated environments to investigate gravitational collapse and implications for the star formation rate \citep{shu77, foster93, hetem93, klessen00, dale05, gavagnin17}. Follow-up studies started to investigate connections to the surrounding ISM with idealised gas replenishing scenarios such as colliding flows \citep{heitsch05, vazquezsemadeni07, vazquezsemadeni10} or cloud-cloud collisions \citep{Whitworth1994, Inoue2013, Dobbs2015, Balfour2015}. In galactic-scale simulations \citep{deavillez05, slyz05, joung06, hill12, kim13, hennebelle14, Smith2014, walch15, dobbs15, girichidis16} the statistical properties of MCs are analysed and the global importance of individual feedback processes estimated \citep{girichidis16, padoan16, gatto17, peters17, Padoan2017, KO2018}. Recent progress in computational performance enables us to simulate a galactic-scale environment and simultaneously increase the spatial and time resolution in forming molecular clouds. This technique is referred to as a \textit{zoom-in} simulation \citep{clark12b, bonnell13, smith14, dobbs15, butler17, ibanezmejia17, kuffmeier17, pettitt17, seifried17, Nordlund2017}. The advantage is that large scale influences (e.g. SN shocks) are propagated down to MC-scales and cloud formation can be studied in a self-consistently evolved environment.

The impact of stellar feedback strongly depends on the mass of the star, hence the UV-luminosity \citep{geen18}, and its environment. A massive star with $M_{\ast} \approx$ 23 M$_{\odot}$ emits a factor of $\sim$100 less energy in a wind than it releases in radiative energy \citep{matzner02} but higher/lower mass stars have stronger/weaker winds relative to radiation. Furthermore, stellar winds are inefficiently coupled to dense environment \citep{haid17}. Therefore, in massive MCs ($M \gtrsim$ 10$^{5}$ M$_{\odot}$), the impact of stellar winds seems negligible \citep{dale14, geen15, ngoumou15, howard17}. However, in low-mass MCs ($M \approx$ 10$^{4}$ M$_{\odot}$), winds are able to reshape the clouds, ablate dense material, and even drive gas out of the clouds through low density channels \citep{rogers13}. Winds are also more important than radiation if the environment of the massive star is already warm or hot, because in this case the radiation does not couple to the surrounding ISM and the radiative energy cannot be deposited in the gas \citep[i.e. low coupling efficiency; ][]{haid17}. Ionizing radiation also struggles to impact bound, massive MCs \citep{dale12, dale13}, while clouds with the sound speed of the photo-ionized gas being similar to the escape velocity can be dispersed completely within a few Myr \citep{walch12}. In any case, both processes shape the environment for the final SN explosions to leak out, thereby dispersing the clouds effectively \citep{harperclark09, pittard13, rosen14, gatto17, naab17, peters17, wareing17}.

The observed star formation in MCs is low with only a few percent of gas that is converted into stars during one free-fall time \citep{zuckerman74, evans09, murray11a}. This inefficiency suggests that processes inside a cloud oppose the gravitational collapse. Stellar feedback is discussed to be an internal driver of supersonic turbulence with a velocity dispersion of a few km s$^{-1}$ \citep{maclow03, maclow04, mellema06, walch12}. However, numerical simulations fail to reproduce this low level of star formation \citep{klessen00, vazquezsemadeni03, dale14}.

Therefore, two aspects of the interaction of stellar feedback with the MC environment remain a matter of discussion. Is star formation limited to the low observed values of a few percent as a consequence of internal feedback processes?  What is the role of MC substructure and filling factor on the coupling efficiencies of stellar winds and ionizing radiation \citep{haid17}?

To address this questions, we present three-dimensional, radiation-hydrodynamic adaptive mesh refinement (AMR) simulations of MCs as part of a SN-driven, multi-phase ISM in a piece of a galactic disc \citep[within the SILCC project;][]{walch15, girichidis16}. We apply a zoom-in technique to follow the formation and evolution of two MCs with total gas masses of a few 10$^{4}$ M$_{\odot}$ with an effective resolution of 0.122 pc \citep{seifried17}. Sink particles are integrated by our novel predictor-corrector scheme (Dinnbier et al., in prep.). With a model of star cluster formation within sink particles, we couple ionizing radiation to the ambient medium \citep{haid17}. The radiation is treated by the novel, tree-based radiative transfer scheme \textsc{TreeRay} (W\"unsch et al. 2018, in prep.) based on the tree solver for gravity and diffuse radiation implemented in {\sc Flash} \citep{wunsch17}. For now, we neglect stellar winds, as their contribution in the early, dense phase of MCs is likely subordinate to ionizing radiation \citep{dale14, haid17}. We focus on the interplay of ionizing radiation and the particular MC morphology and star formation efficiency.

The paper is organized as follows. In Section \ref{sec-numerical}, we present the numerical method. In Section \ref{sec-setup}, we give an overview of the simulation setup. We discuss the morphological impact of ionizing radiation in Section \ref{sec-morphology}. We depict the effect of radiative feedback on the environment around the stellar component (Section \ref{sec-environment}). The evolution of global cloud properties is shown in Section \ref{sec-difference} and the differences between the two clouds are discussed in Section \ref{sec-difference-mc1mc2}. Finally, we conclude in Section \ref{sec-summary}.

\section{Numerical method}
\label{sec-numerical}

We use the three-dimensional AMR magneto-hydrodynamics code (MHD) \textsc{FLASH} 4 \citep{fryxell00,dubey08} with the directionally split, Bouchut HLL5R solver \citep{bouchut07, bouchut10, waagan09, waagan11} including self-gravity, a chemical network to follow molecule formation and dissociation, the novel radiative transfer module \textsc{TreeRay}, sink particles, and the stellar evolution of massive stars.

\subsection{Sink particles}
\label{sec-numerical-sink}
Sink particles represent the unresolved formation of stars or clusters by gravitational collapse. In the simulations, we use a new particle module (Dinnbier et al., in prep.) which uses a Hermite predictor-corrector integrator and is coupled to the Barnes-Hut tree \citep{wunsch17}. The sink formation and accretion criteria are the same as in \citet{federrath10}. In this work, sink particles represent star clusters (hereafter also simply called sinks) within which multiple massive stars (hereafter also stars) can form. For further information on the cluster sink implementation we refer to \citet{gatto17}.

A sink particle can only be formed in a computational cell and followed through the computational domain if the harbouring cell lives on the highest refinement level (smallest spatial resolution) in the AMR hierarchy. The accretion radius $r_{\text{accr}}$ is set to $r_{\text{accr}}$ = $2.5\times\Delta x$ = 0.31 pc. We further demand that the gas within $r_{\text{accr}}$  is Jeans unstable, is in a converging flow and represents a local gravitational potential minimum \citep{federrath10}. Under the assumption of an isothermal gas with a temperature $T$ = 100 K, we derive the density threshold above which sinks can form, $\rho_{\text{si}}$ =  1.1$\times$10$^{-20}$ g cm$^{-3}$ following the Jeans criterion:
\begin{equation}
\rho_{\text{si}} = \frac{\pi k_{\text{B}}}{m_{\text{P}}G} \frac{T}{4r_{\text{accr}}^{2}}
\end{equation}
where $k_{\text{B}}$ is the Boltzmann constant and $G$ the gravitational constant \citep{truelove97, gatto17}.

Sink particles accrete gas. A fraction of the accreted gas is turned into massive stars by means of the star cluster subgrid model. Assuming a Kroupa stellar initial mass function (IMF), one new massive star (9 M$_{\odot}$ $\leq$ $M_{\ast}$ $\leq$ 120 M$_{\odot}$) is randomly sampled for every 120 M$_{\odot}$ accreted on a sink \citep{kroupa01}. We assume the Salpeter slope of -2.35 in the high mass regime of the IMF \citep{salpeter55}. Each sink with a mass $M_{\text{si}}$ can contain $N_{\ast}$ stars with individual initial masses $M_{\ast}$ and individual stellar evolutions tracks (see \citealt{ekstroem12}, \citealt{gatto17}, \citealt{peters17} and references therein). We refer to the number of massive stars, $N_{\ast}$, in a sink as the \textit{active stellar component}, $M_{\ast, \text{tot}}$, with $M_{\ast,\text{tot}} = \sum^{N_{\ast}}_{i=1} M_{\ast, \text{i}}$. The residual gas is converted into low-mass stars, which are not recorded individually as they currently provide no feedback to the surrounding medium.

Each sink particle is subject to the gravitational attraction of the gas and the other sink particles. Their trajectories are integrated by a predictor-corrector scheme, which is inspired by the two nested fourth-order Hermite predictor-corrector integrators used in the \textsc{Nbody6} code \citep{Makino1991,Makino1992, aarseth99, Aarseth2003}. Here, the outer (regular) integrator takes into account the slowly varying force due to the gas, while the inner (irregular) integrator takes into account the fast varying force due to the other sink particles. It is an anlogue to the Ahmad-Cohen scheme \citep{Ahmad1973} where the division to regular and irregular forces is based on the kind of interaction (gas or sink particles) instead of physical proximity. The regular time-step corresponds to the hydrodynamical time-step, while the irregular time-step $\Delta t_{\text{irr}}$ is calculated according to the standard formula \citep{Aarseth2003}
\begin{equation}
\Delta t_{\text{irr}}  = \left( \frac{\eta( |\textbf{a}||\ddot{\textbf{a}}| + |\dot{\textbf{a}}|^2)}
{|\dot{\textbf{a}}||\dddot{\textbf{a}}| + |\ddot{\textbf{a}}|^2} \right)^{1/2},
\label{etstep}
\end{equation}
and then quantised to bins differing by factor of 2 in time. We set the constant for integration $\eta$ to be $\eta = 0.01$. The quantities $\textbf{a}$,  $\dot{\textbf{a}}$, $\ddot{\textbf{a}}$ and $\dddot{\textbf{a}}$ are the acceleration and the higher time derivatives acting on the particle due to the other sink particles. The scheme uses the softening kernel described in \citet{Monaghan1985} with softening length corresponding to $2.5\times\Delta x$ = 0.31 pc at the highest refinement level. Likewise, gas is attracted by sink particles, which are placed to the tree to facilitate the force evaluation. We present the detailed description of the sink particle integrator as well as numerical tests in Dinnbier et al. (in prep.).

\subsection{Ionizing radiation and radiative heating}
\label{sec-numerical-radiation}
\begin{figure}
\includegraphics[width=0.5\textwidth]{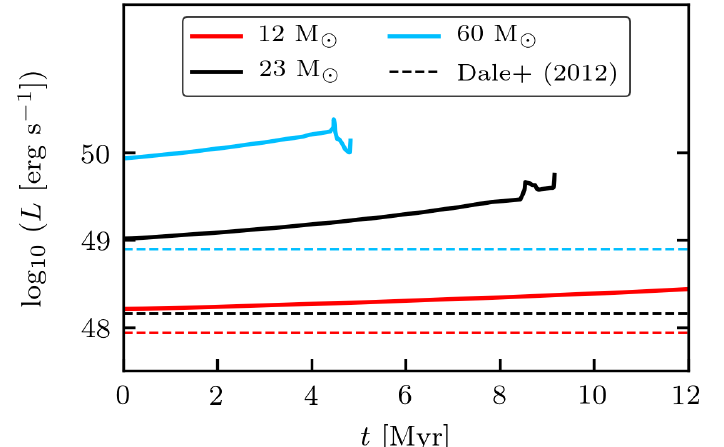} 
\caption{Time evolution of the radiative luminosities $L_{\text{RAD}}$ of stars with $M_{\ast}$ = 12 (red), 23 (black) and 60 M$_{\odot}$ (blue). The dashed lines show the values used in the simulations of \citet{dale12}. Note that the end of the evolution of the 12 M$_{\odot}$ is not shown here.}
\label{fig-evolution-tracks}
\end{figure}

The transfer of ionizing radiation is calculated by a new module for the \textsc{Flash} code called \textsc{TreeRay}. It is an extension of the \textsc{Flash} tree solver described in \citet{wunsch17}. \textsc{TreeRay} uses the octal-tree data structure constructed and updated at each time step by the tree solver and shares it with the \textsc{Gravity} (calculates gas self-gravity, see \citealt{wunsch17}), \textsc{Optical-Depth} (calculates the optical depth and parameters for the total, H$_{2}$ and CO shielding, see \citealt{walch15, wunsch17}), and \textsc{EUV} modules. The latter is the new module which calculates the local flux of ionizing radiation. Here, we only give basic information about \textsc{TreeRay}; a detailed description alongside with accuracy and performance tests will be presented in W\"unsch et al. (2018, in prep). \textsc{TreeRay} has already been benchmarked in \citet{bisbas15} and applied in homogeneous media \citep{haid17}.

Each node of the octal-tree represents a cuboidal collection of grid cells and stores the total gas mass contained in it, masses of individual chemical species and the position of the mass centre. In addition to that, \textsc{TreeRay} stores for each node the total amount of the radiation luminosity generated inside the node, the radiation energy flux passing through the node, and the node volume. Before the tree is traversed for each grid cell (called {\em target cell}), a system of $N_{\text{pix}}$ rays pointing from the target cell to different directions is constructed. The directions are determined by the \textsc{HEALPIX} algorithm \citep{gorski05}, which tessellates the unit sphere into elements of equal spatial angle. We use $N_{\text{pix}}$ = 48. Each ray is then divided into segments with lengths increasing linearly with the distance from the target cell. In this way, the segment lengths correspond approximately to the sizes of the nodes interacting with the target cell during the tree walk if the Barnes-Hut (BH) criterion for node acceptance is used. Here, we use the BH criterion with an opening angle of $\theta_{\rm lim} = 0.5$. When the tree is traversed, node densities, radiation luminosities, and energy fluxes are mapped onto the ray according to the node and the volume belonging to the ray segment.

Finally, after the tree walk, the one-dimensional radiative transport equation is solved using the \textit{On-the-Spot approximation} along each ray using the case B recombination coefficient $\alpha_{\text{B}}$ with the temperature dependence in the range of $T$ = [5000, 20000] K given by \citep{draine11}
\begin{equation}
\label{eq-alphab}
\alpha_{\text{B}} = 2.56\times10^{-13}  \text{cm}^{3}\ \text{s}^{-1} \left(\frac{T}{10^{4}\ \text{K}}\right)^{-0.83}. 
\end{equation} 
The radiative transfer equation along the ray towards the target cell is given by
\begin{equation}
F_{0} = \sum_{i=N-1}^{0} \left[\frac{\epsilon_{\text{i}}}{4\pi r_{\text{i}}^{2}} - \sum_{j=i+1}^{N}\alpha_{\text{B}}\frac{F_{\text{i,j}}^{2}}{F_{\text{tot,i}}} \text{d}V_{\text{ij}} \right]
\end{equation}
where $F_{0}$ is the received flux in the target cell, $N$ is the number of segments along a ray, $\epsilon_{\rm i}$ is the emission coefficient in segment $i$, $F_{\text{tot,i}}$ the total flux coming into segment $i$, $F_{\text{i,j}}$ the source in a segment if a source exists, $r_{i}$ the distance from the segment to the target cell, and $\text{d}V_{\text{ij}}$ the volume of the segment. As the radiation flux passing through a given segment from different directions has to be taken into account, the solution has to be searched for iteratively, repeating the whole process of tree construction, tree walk and solving the radiation transport equation until the maximum relative error drops below 0.01. To speed up convergence, we use the result of the previous hydrodynamic time-step as the radiation field typically changes only slightly between times-steps, in most cases only one or two iterations are needed in each time step.

We use the prescription given in \citet{gatto17} and \citet{peters17} to simulate the evolution, and in particular the radiative energy output, of massive stars using the Geneva stellar tracks from the zero-age main sequence to the Wolf-Rayet phase \citep{kudritzki00, markova04, markova08, puls08,ekstroem12}. An initial proto-stellar phase is not included. The corresponding time evolution of the radiative luminosity $L$ is shown in Fig. \ref{fig-evolution-tracks} for three stars with $M_{*}$ = 12 (red), 23 (black), and 60 M$_{\odot}$ (blue). For the later discussion, we include dashed horizontal lines that correspond to the luminosities used in \citet{dale12}, \citet{dale14} based on stellar models of \citet{diazmiller98}.

The aforementioned stellar tracks provide the time-dependent number of Lyman continuum photons, $\dot{N}_{\text{Lyc}}$, and the effective stellar temperature \citep{peters17}. In \textsc{TreeRay}, this information is processed to get the average excess photon energy, $E_{\bar{\nu} - \nu_{T}}$ between $\nu_{\text{T}}$ = 13.6 eV $h^{-1}$ and the average photon frequency $\bar{\nu}$, by assuming a black-body spectrum for each star and integrating it in the Lyman continuum \citep{rybicki04}. Note that, since we only consider the radiative transfer in a single energy band (all photons in the Lyman continuum), we do not distinguish between the direct ionization of H and H$_{2}$, as necessary for detailed models of photon-dominated regions \citep{rollig07,baczynski15}.

We calculate the heating rate, $\Gamma_{\text{ih}}$, in the ionization-recombination equilibrium with \citep{tielens05}
\begin{equation}
\Gamma_{\text{ih}} = F_{\text{ph}} \sigma E_{\bar{\nu} - \nu_{T}} = n_{\text{H}}^{2} \alpha_{\text{B}}  h \left(\bar{\nu} - \nu_{\text{T}} \right) ,
\label{eq-heatingrate}
\end{equation}
where $F_{\text{ph}}$ is the photon flux, $\sigma$ the hydrogen photo-ionization cross-section, $n_{\text{H}}$ the hydrogen number density, and $h$ is the Planck constant. The ionization heating rate and number of ionizing photons are provided to the \textsc{Chemistry} module (see Section \ref{sec-numerical-chemistry}), where the temperature is self-consistently increased by balancing heating and cooling processes, the mean hydrogen ionization state is updated using the given photo-ionization rate \citep{haid17} and CO is dissociated. In ionization-recombination equilibrium, an H\textsc{ii} region develops around the sink particle with interior temperatures between $\sim$7000~--~9000 K. In homogeneous media, this is well explained by the ionization of the Str\"omgren sphere followed by the \textit{Spitzer} expansion \citep{stromgren39, spitzer78, hosokawa06}. However, the equilibrium temperature strongly depends on the density of the ionized gas within the H\textsc{ii} region and can be significantly lower in young, embedded H\textsc{ii} regions, which are still quite dense (see Section \ref{sec-environment-evolution }) .

\subsection{Gas cooling, heating and chemistry}
\label{sec-numerical-chemistry}

We include a simple chemical network, which is explained in detail in \cite{walch15}. It is based on \citet{glover07b, glover07a, glover10} and \citet{nelson97} to follow the abundances of seven chemical species: molecular, atomic and ionized hydrogen as well as carbon monoxide, ionized carbon, atomic oxygen and free electrons (H$_{2}$, H, H$^{+}$, CO, C$^{+}$, O, e$^{-}$). The gas has solar metallicity \citep{sembach00} with fixed elemental abundances of carbon, oxygen and silicon ($x_{\text{C}}$ = 1.14 $\times$ 10$^{-4}$, $x_{\text{O}}$ = 3.16 $\times$ 10$^{-4}$, $x_{\text{Si}}$ = 1.5 $\times$ 10$^{-5}$) and the dust-to-gas mass ratio is set to 0.01. We include a background interstellar radiation field (ISRF) of homogeneous strength G$_{0}$ = 1.7 \citep{habing68, draine78}. So far, \textsc{TreeRay} does not treat the FUV regime. The effect of radiation in the FUV energy band will be discussed in a follow-up paper. For the cloud dynamics, we still expect photoionization to be the dominant process  \citep{peters10, walch12, baczynski15} with typical temperatures around $\sim$8000~K almost independently of gas densities. FUV radiation is considered to be important in photo-dissociation regions which are forming ahead of the ionization shock fronts. As we show in Section \ref{sec-environment-evolution }, a hypothetical FUV field of 1000 $\times$ G$_{0}$ increases the gas temperature in such dense ($\approx$10$^{-21}$ g cm$^{-3}$) photo-dissociation regions to a few 100~K at most. Therefore, the predicted dynamical effect resulting from the FUV heating is considered to be negligible in respect to the EUV heating.

The ISRF is attenuated in shielded regions depending on the column densities of total gas, H$_{2}$, and CO. Thus, we consider dust shielding and molecular (self-) shielding for H$_{2}$ and CO \citep{glover10} by calculating the shielding coefficients with the \textsc{TreeRay Optical-Depth} module \citep{wunsch17}. From the effective column density in each cell the visual extinction, $A_{\text{v}}$, is calculated by 
\begin{equation}
A_{\text{v}} = \frac{N_{\text{H}}}{1.8 \times 10^{21}\;{\rm cm}^{2}},
\end{equation} 
where the total gas column density $N_{\text{H}}$ is given by  $N_{\text{H}}$~=~$\Sigma_{\text{H}}/(\mu m_{\text{p}})$ where $\Sigma_{\text{H}}$ is the surface density, $\mu$ the mean molecular weight and $m_{\text{p}}$ the proton mass. 

For gas with temperatures above 10$^4$ K we model the cooling rates according to \citet{gnat12} in collisional ionization equilibrium. Non-equilibrium cooling (also for Lyman $\alpha$) is followed at lower temperatures through the chemical network. Within the H\textsc{ii} region, we neglect both C$^{+}$ and O cooling because these species are predominantly in a higher ionization state. Heating rates include the photoelectric effect, cosmic ray ionization with a rate of $\xi$ = 3$\times$10$^{-17}$ s$^{-1}$, X-ray ionization by \cite{wolfire95b}, and photo-ionization heating (see Section \ref{sec-numerical-radiation}).

\section{Simulation setup}
\label{sec-setup}
\subsection{The SILCC simulation}
\label{sec-setup-silcc}

The SILCC simulation \citep{walch15, girichidis16} is the basic setup and is used to self-consistently study the evolution of the SN-driven multi-phase ISM. The computational domain with an extent of 500~pc~x~500~pc~$\pm$5~kpc has a disc midplane with galactic properties at low red-shift similar to the solar neighbourhood. The boundary conditions for the gas are periodic in $x$- and $y$- direction and outflow in $z$-direction. For gravity the boundary conditions are periodic in $x$- and $y$- direction and isolated in $z$-direction (see \citealt{wunsch17} for mixed gravity boundary conditions. The base grid resolution, denoted as l$_{\text{ref}}$ = 5 in the following, is $\Delta x$ = 3.9~pc.

Initially, the disc has a gas surface density of  $\Sigma_{\text{Gas}}$ = 10 M$_{\odot}$~pc$^{-2}$ and the density profile follows a Gaussian distribution in the vertical direction
\begin{equation}
\rho(z) = \rho_{0}\exp{ \left[- \frac{1}{2}\left( \frac{z}{h_{\text{z}}}\right)^{2}\right]},
\end{equation} 
where the scale height of the gas $h_{\text{z}}$ = 30 pc and the mid-plane density $\rho_{0}$ = 9 $\times$ 10$^{-24}$ g cm$^{-3}$. The initial temperature of the gas near the mid-plane is 4500 K and the disc is made up of H and C$^{+}$. The density is floored to 10$^{-28}$ g~cm$^{-3}$ with a temperature of 4 $\times$ 10$^{8}$ K in the gas at high altitudes above and below the galactic plane. 

The simulation includes a static background potential to model the old and inactive stellar component in the disc, which is modelled as an isothermal sheet with a stellar surface density $\Sigma_{\ast}$ = 30 M$_{\odot}$ pc$^{-2}$ and a scale height of 100 pc \citep{spitzer42}. This static potential is added to the gravitational potential of the self-gravitating gas that is calculated in every time-step. 

For the first $t_{\text{ZI}}$ = 11.9 Myr from the start of the simulations, the development of a multi-phase ISM is driven by SN explosions. Therefore, we inject SNe at a fixed rate of 15 Myr$^{-1}$. The SN rate bases on the Kennicutt-Schmidt relation \citep{schmidt59, kennicutt98} and a standard initial mass function for $\Sigma_{\text{Gas}}$. We use mixed SN driving, where SNe explode in density peaks and at random positions by an equal share (1:1 ratio). In $z$-direction, the positioning of the random SNe is weighted with a Gaussian distribution with a scale height of 50 pc for (see \citealt{walch15} and \citealt{girichidis16} for more details).

A single SN injects 10$^{51}$ erg of energy. Whether the energy is injected in the form of internal energy or momentum depends on the ability to resolve the Sedov-Taylor radius. The spherical injection region has a minimum radius of four grid cells. In case the density is low, the Sedov-Taylor radius is resolved and the energy thermally injected. If it is unresolved, the temperature in the region is raised to $T$ = 10$^{4}$ K and the SN bubble is momentum-driven \citep{blondin98, gatto15, walch15, haid16}.

\subsection{Initial conditions for this work: the zoom-in simulations}
\label{sec-setup-zoomin}

We refer to the re-simulation of selected clouds with a higher spatial resolution as \textit{zoom-in}. Two different molecular clouds, MC$_{1}$ and MC$_{2}$, are selected from the SILCC-ZOOM simulation \citep{seifried17}. The selected domains (see Table \ref{tab-setup1}) are traced back in time to properly model the formation process from the beginning. The zoom-in starts at $t_{\text{ZI}}$ = 11.9 Myr  where SN driving is suspended and the typical number densities in the selected zoom-in regions do not exceed some 10 cm$^{-3}$.

During the zoom-in simulation, the resolution is gradually increased in both MCs. Starting from the SILCC base grid resolution of $\Delta x$ = 3.9 pc (l$_{\text{ref}}$ = 5), we allow adaptive refinement down to  $\Delta x$= 0.122 pc (l$_{\text{ref}}$ = 10). Two refinement criteria are used. The refinement on the second derivative of gas densities, which picks up density fluctuations, is limited to a maximum refinement of 0.5 pc. Further refinement to the smallest $\Delta x$ depends on the local Jeans length, $L_{\text{Jeans}}$, which is computed for each cell. We require that $L_{\text{Jeans}}$ is resolved with at least 16 cells in each spatial dimension, otherwise we refine. 

The zoom-in is not carried out in a single time-step. Starting from the base grid resolution, we increase the refinement step-by-step and require about 200 time steps in between two steps. On the one hand, this choice allows the relaxation of the gas to prevent filamentary grid artefacts, which appear in case of an instantaneous zoom-in \citep{seifried17}. On the other hand, it avoids the formation of large-scale, rotating, disc-like structures in case of a slower zoom-in in which case compressive motions are dissipated too efficiently. The zoom-in simulation reaches the highest refinement level at $t_{\text{ZE}}$ = 13.2 Myr.

\subsection{Simulation overview}
\label{sec-setup-simulation}

\begin{figure}
\includegraphics[width=0.5\textwidth]{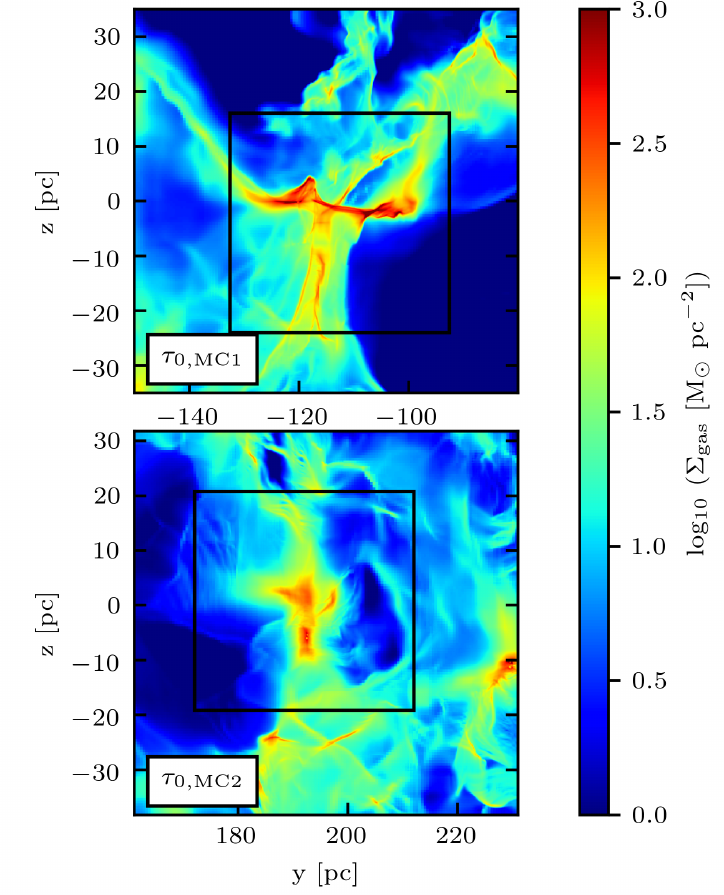}
\caption{Gas column density $\Sigma_{\text{gas}}$ in the $y$-$z$-plane for the total volume V$_{\text{tot}}$ of cloud MC$_{1}$ (top) and MC$_{2}$ (bottom) at the formation time of the first massive star, at $t_{\text{0,MC1}}$ = 13.73 and $t_{\text{0,MC2}}$ = 13.55 Myr. The black frames indicate the central volume V$_{\text{CoV}}$ to be shown in more detail in Fig. \ref{fig-morphology-mc1} and Fig. \ref{fig-morphology-mc2}}.
\label{fig-setup-init}
\end{figure} 

In this work, we continue from the zoom-in simulation at $t_{\text{ZE}}$ = 13.2 Myr and allow for the formation of cluster sink particles. From this time we start two simulations. The reference run, ZI\_NOFB, does not include any stellar feedback. In the second simulation, ZI\_RAD, the forming, active stellar component provides ionizing radiation. 

Note that each simulation contains both molecular clouds, MC$_{1}$ and MC$_{2}$ with similar volumes $V_{\text{tot}}$, in which the zoom-in is enabled, and total gas masses $M_{\text{tot}}$ within 10 percent. Table \ref{tab-setup1} summarizes the initial properties of the clouds. Each cloud develops its own sink evolution, i.e. star formation history. The first sink in MC$_{1}$ forms at $t_{0,\text{MC1}}$ = 13.51 Myr and in MC$_{2}$ at $t_{0,\text{MC2}}$ = 13.40 Myr. As radiation feedback sets in, the clouds start to evolve differently. Therefore, we define six times $\tau_{i}$ relative to $t_0$, which we use for the analysis of the clouds as:
\begin{equation}
\tau_{i} \equiv t_{i} -  t_{0}\ \text{with}\ i \in \left[0,1.0,1.5,2.0,2.5,3.0\right].
\end{equation}
The subscript $i$ indicates the time in Myr after $t_{0}$, i.e. $\tau_{0}$ = 0 Myr refers to $t_0$ and $\tau_{3.0}$ = 3.0 Myr after $t_0$. A second subscript is used to indicate the respective MC.

Fig. \ref{fig-setup-init} shows the total gas column density, $\Sigma_{\text{gas}}$, in the $y$-$z$-plane in MC$_{1}$ (top) and MC$_{2}$ (bottom) at $\tau_{0}$. To obtain $\Sigma_{\text{gas}}$, we integrate the density $\rho$ along the $x$-direction within the volume $V_{\text{tot}}$ (see Table \ref{tab-setup1}). The black frames indicate the central subregions (later referred to as center of volume, CoV) to be shown in more detail in Fig. \ref{fig-morphology-mc1} and Fig. \ref{fig-morphology-mc2} with properties summarized in Table \ref{tab-setup1}.

\begin{table}
\caption{Overview of the total (top, subscript \textit{tot}) and the central part (bottom, subscript \textit{CoV}) of MC$_{1}$ and MC$_{2}$ with the center $c$ in $x$,$y$,$z$ coordinates (second column), the side length $d$ in $x$,$y$,$z$ direction (third column), the gas mass $M$ (forth column), and the escape velocity $v_{\text{esc}}$ of the cloud (last column) at $t_{0,\text{MC1}}$ = 13.51 Myr and $t_{0,\text{MC2}}$ = 13.40 Myr.}
\label{tab-setup1}
\centering
\begin{tabular}{|l|c|c|c|c}
\hline 
\rule[-1ex]{0pt}{2.5ex} Cloud & $c_{\text{tot}}$ [pc] & $d_{\text{tot}}$ [pc] & $M_{\text{tot}}$ [M$_{\odot}$] & $v_{\text{esc}}$ [km s$^{-1}$] \\ 
\hline
\hline  
\rule[-1ex]{0pt}{2.5ex} MC$_{1}$ & 157,-115,0 &  88,87,77 & 1.0 $\times$10$^{5}$& 5.3 \\ 
\hline 
\rule[-1ex]{0pt}{2.5ex} MC$_{2}$ & 45,196,-3 & 87,87,71 & 8.5 $\times$10$^{4}$ & 4.5 \\ 
\hline
\rule[-1ex]{0pt}{2.5ex}  & & & & \\
\rule[-1ex]{0pt}{2.5ex}  & $c_{\text{CoV}}$ [pc] & $d_{\text{CoV}}$ [pc] & $M_{\text{CoV}}$ [M$_{\odot}$]  & $v_{\text{esc}}$ [km s$^{-1}$] \\ 
\hline
\hline  
\rule[-1ex]{0pt}{2.5ex} MC$_{1}$ & 127,-112,-4 &  40,40,40  & 4.0 $\times$10$^{4}$ & 7.1 \\ 
\hline 
\rule[-1ex]{0pt}{2.5ex} MC$_{2}$ & 55,192,1 & 40,40,40 & 2.7 $\times$10$^{4}$ & 5.2 \\
\hline
\end{tabular} 
\end{table}


\section{The morphology of the molecular clouds}
\label{sec-morphology}
\begin{figure*}
\begin{center}
\includegraphics[width=1\textwidth]{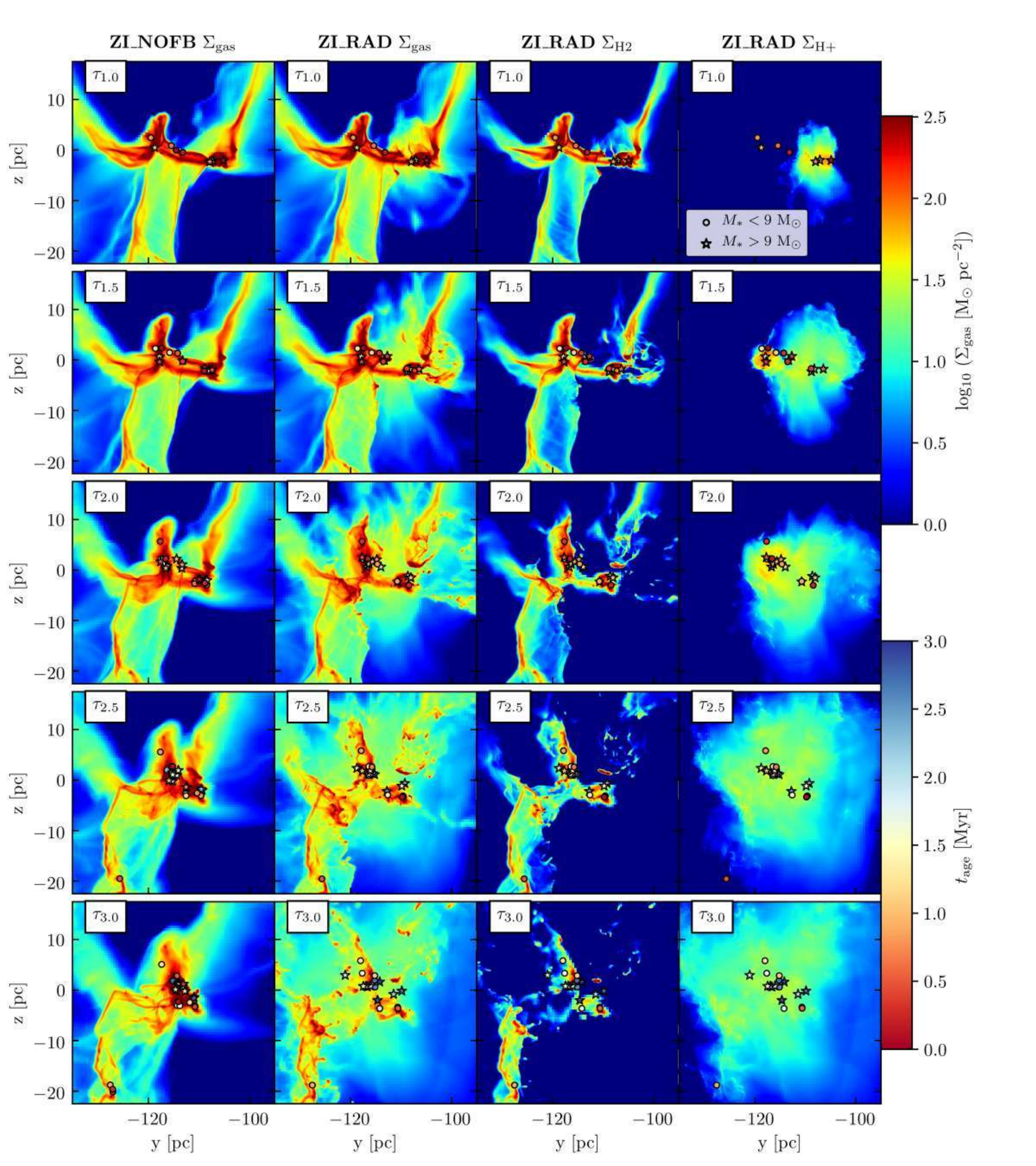}
\caption{Time evolution of the central volume $V_{\text{CoV}}$ of size (40 pc)$^3$ (see Fig. \ref{fig-setup-init} and Table \ref{tab-setup1}) of MC$_{1}$ in the simulations ZI\_NOFB (left) and ZI\_RAD (second to forth), respectively, at times $\tau_{1.0}$ to $\tau_{3.0}$ (from top to bottom). The first two columns show the gas column density $\Sigma_{\text{gas}}$ in the $y$-$z$-plane. The third and forth show the H$_{2}$ and H$^{+}$ column densities. Circles indicate sink particles without an active stellar component, i.e. without massive stars. Star-shaped markers are cluster sink particles with active stellar feedback. The color of the markers indicate their age.}
\label{fig-morphology-mc1}
\end{center}
\end{figure*} 

\begin{figure*}
\includegraphics[width=1\textwidth]{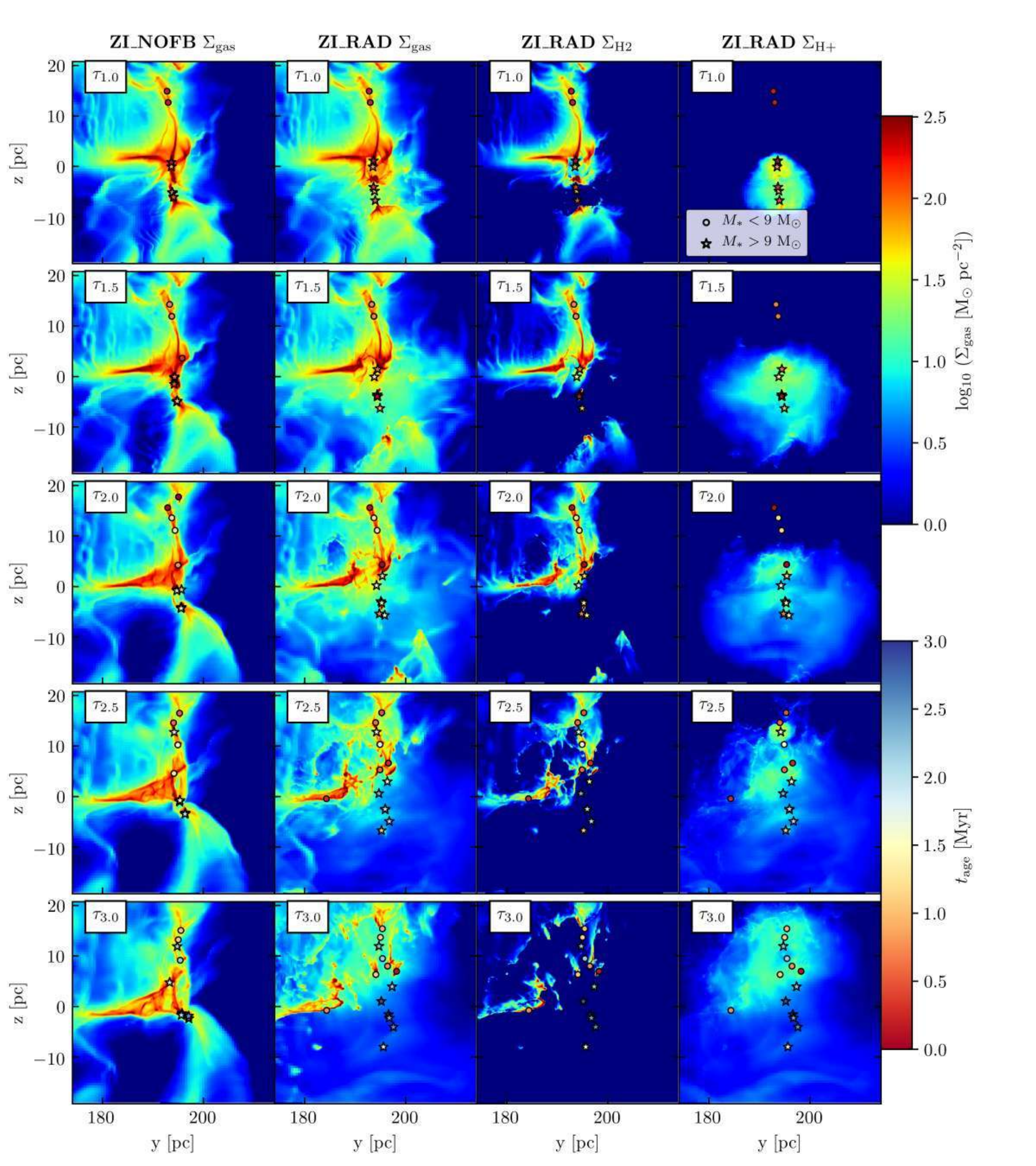}
\caption{Same figure as in Fig. \ref{fig-morphology-mc1} but for MC$_{2}$.}
\label{fig-morphology-mc2}
\end{figure*} 
The masses and volumes of MC$_{1}$ and MC$_{2}$ are comparable (see Table \ref{tab-setup1}). However, their formation out of the turbulent, multi-phase ISM leads to different morphologies (see Fig. \ref{fig-setup-init}). MC$_{1}$ contains a highly collimated, dense ($\Sigma_{\text{gas}} \approx$~500~M$_{\odot}$ pc$^{-2}$), T-shaped filament, where the bar is one horizontal structure with extended ends. The vertical trunk is divided into two, roughly parallel substructures. Each of the dense filaments is surrounded by an 'envelope' with intermediate column densities ($\Sigma_{\text{gas}} \approx$ 5~--~50 M$_{\odot}$ pc$^{-2}$). To the bottom, right a low column density ($\Sigma_{\text{gas}}  \approx$ 0.05~--~0.5 M$_{\odot}$ pc$^{-2}$) cavity is situated, which originates from a previous SN explosion outside of the cloud. In MC$_{2}$ the main filamentary structure is vertically elongated and less condensed with a central, hub-like condensation. Qualitatively, the surface density maps of MC1 and MC2 span the same dynamic range.

In Fig. \ref{fig-morphology-mc1} and Fig. \ref{fig-morphology-mc2} we show the time evolution (from top to bottom) of the column densities of MC$_{1}$ and MC$_{2}$ without (leftmost column) and with radiative feedback (2$^{\rm nd}$ column). Sink particles without and with active stellar components are indicated with circles and stars, respectively with their age indicated by a second color scheme ranging from 0 to 3 Myr. The third and 4$^{\rm th}$ column show the column densities of molecular hydrogen and ionized hydrogen for the runs with radiation. Here, we only show the central (40~pc x 40~pc) subregion of the clouds (see Table \ref{tab-setup1}, bottom, subscript 'CoV') and therefore the column densities are obtained from the integration along the $x$-direction over the corresponding 40 pc, henceforth $V_{\text{CoV}}$. By comparing the maximum values of $\Sigma_{\text{gas}}$ in Fig.~\ref{fig-setup-init} with Fig.~\ref{fig-morphology-mc1} and Fig.~\ref{fig-morphology-mc2}, one can see that only lower density gas from the fore- and background has been cut. 

Without feedback (runs ZI\_NOFB), gravity is further condensing the initial structures while the lower column density gas surrounding the main filaments is accreted. In MC$_{2}$ the gas is gravitationally collapsing but the global structure of the cloud does not change significantly and is still recognizable at $\tau_{3.0}$. In both clouds, sink formation occurs in the densest filament(s) and its debris. 

Radiative feedback (runs ZI\_RAD), does not significantly alter the global dynamics of MC$_{1}$ during the first 2 Myr but more filamentary substructures appear while the existing substructures seem to be locally supported against gravitational collapse. The dense regions are puffed up by the expanding radiative shocks. Multiple radiation driven, partly or fully embedded bubbles develop (see $\Sigma_{\text{H+}}$ in the right columns of Fig. \ref{fig-morphology-mc1} and Fig.~\ref{fig-morphology-mc2}). Some active sinks do not form a noticeable bubble of ionized hydrogen, in particular if the contained massive stars that are less massive than 20 M$_{\odot}$. During the last Myr, the clouds decompose and a variety of filamentary substructures evolve into all directions. The envelope is widened and heated gas is expelled into the cavity. Star formation takes place in the bar and its remnants but primarily in the central dense clump (compare to $\Sigma_{\text{H2}}$ in the third column of Fig. \ref{fig-morphology-mc1}). In MC$_{2}$, the bottom half of the cloud forms massive stars quickly, while the upper half forms only low-mass and hence inactive sink particles. Feedback from the bottom half disrupts the cloud into an upper, crescent-shaped filament (compare to $\Sigma_{\text{H2}}$ in Fig. \ref{fig-morphology-mc2}, third column) and some left-over, dispersed gas below (see $\Sigma_{\text{H+}}$ at $\tau_{3.0}$ in Fig. \ref{fig-morphology-mc2}, forth column) . At later time, the emerging feedback triggers a second generation (y = 195 pc, z = 5 pc) of massive stars in the upper part. The low-density envelope is replenished with expelled gas.

MC$_{1}$ and MC$_{2}$ with radiative feedback show a significant difference in morphology. The first cloud evolves into one massive structure with multiple embedded H\textsc{ii} regions and is surrounded by a low density envelope, which is only slowly evolving. The central structure hosts almost all stars and star formation continues. The second cloud is partly destroyed by a rapidly forming first generation of stars in the lower cloud filament. A new generation of stars is triggered in the upper part of the central subregion. That demonstrates that not only the mass (which is roughly similar for both clouds) but also the morphology prior to stellar feedback influences its impact. We investigate this further in Section \ref{sec-difference}.

\section{Cloud environments with massive stars}
\label{sec-environment}

\begin{figure}
\includegraphics[width=0.5\textwidth]{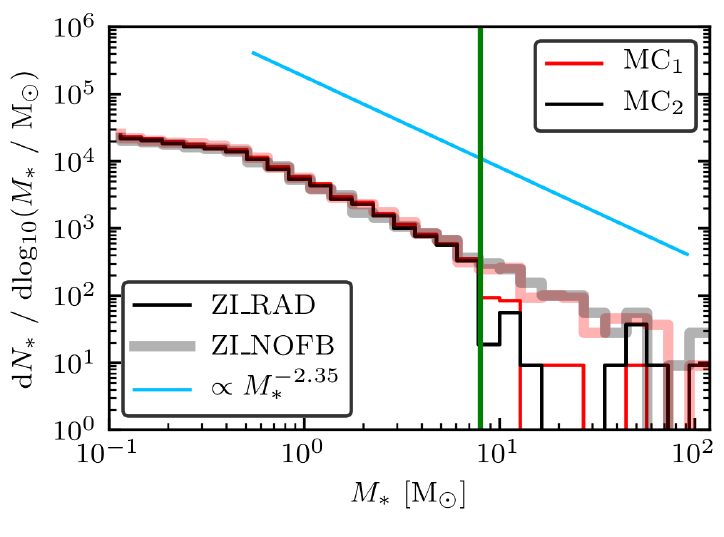}
\caption{Initial mass function in MC$_{1}$ (red) and MC$_{2}$ (black) for simulation ZI\_NOFB (thick) and ZI\_RAD (thin) at $\tau_{3.0}$. The blue line indicates the Salpeter slope proportional to $M_{\ast}^{-2.35}$ \citep{salpeter55}. With radiation within this first 3 Myr, 31 and 23 massive stars form within MC$_{1}$ and MC$_{2}$, respectively. Hence, the high mass range ($M_{\ast}$ $>$ 9 $M_{\odot}$) suffer from low number statistics. However, sampling low mass stars from the residual sink mass shows that the underlying Kroupa IMF is well represented \citep{kroupa01}. The green, vertical line indicates the boarder between the low-mass and high-mass regime at 9 M$_{\odot}$.}
\label{fig-environment-imf}
\end{figure} 

In runs ZI\_NOFB, the total numbers of sinks in MC$_{1}$ and MC$_{2}$ at time $\tau_{3.0}$ is 39 and 19 with masses of $\sim$1.8$\times$10$^{4}$ and 1.5$\times$10$^{4}$ M$_{\odot}$, respectively. In ZI\_RAD, the two clouds host 31 sinks with $\sim$5900 M$_{\odot}$ and 23 sinks with $\sim$3300 M$_{\odot}$, respectively (see top panel of Fig. \ref{fig-appendix-sistmass}). Hence in MC$_{1}$ more sinks with a smaller average mass per sink particle are formed than in MC$_{2}$. The different fragmentation properties of the two clouds is caused by the different cloud substructure. In MC$_{2}$, the total number of sinks is slightly increased by radiative feedback, although the mass in sinks is dramatically reduced. This shows that radiative feedback may regulate star formation and, at the same time, trigger star formation. A more detailed investigation of triggered star formation is postponed to a follow-up paper.

The IMF for massive stars in MC$_{1}$ (red) and MC$_{2}$ (black) is shown in Fig. \ref{fig-environment-imf} for the simulations ZI\_NOFB (thick, transparent lines) and ZI\_RAD (thin, opaque lines) at time $\tau_{3.0}$. The blue line indicates the Salpeter slope of the IMF in the high mass range proportional to $M_{\ast}^{-2.35}$ \citep{salpeter55}. In the runs ZI\_RAD, 31 and 23 massive stars form with $M_{\ast, \text{tot}} \approx$  830 and 480 M$_{\odot}$ in the total volume of MC$_{1}$ and MC$_{2}$, respectively. Within the period of 3 Myr, a small number of massive stars is forming, which leaves the high-mass end of the IMF under-sampled. Hypothetical sampling of low-mass stars ($M_{*} <$ 9 M$_{\odot}$) from the residual sink mass results in a well-represented low-mass end of the Kroupa IMF. The corresponding formation history of the massive stars is shown in Fig.~\ref{fig-appendix-lum}.

\begin{figure}
\includegraphics[width=0.5\textwidth]{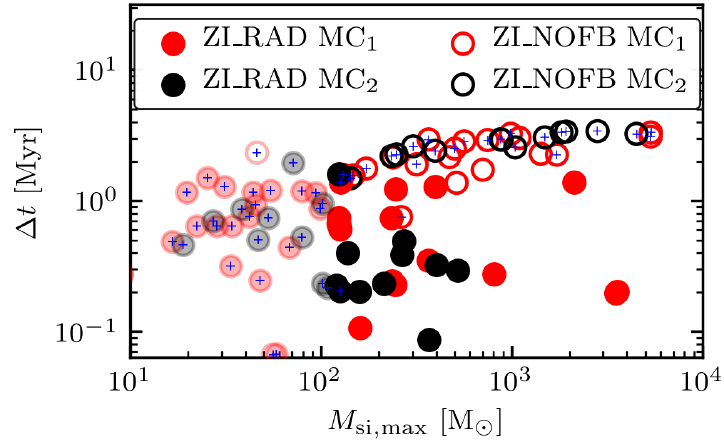}
\caption{Accretion time $\Delta t_{\text{acc}}$ counted from the formation time of the sink as a function of the maximum sink mass $M_{\text{si, max}}$ for the simulations ZI\_RAD (full markers) and ZI\_NOFB (open markers) within $V_{\text{CoV}}$ of MC$_{1}$ (red) and MC$_{2}$ (black). Transparent markers indicate that the sink mass does not exceed the 120 M$_{\odot}$ high mass star forming threshold. Blue crosses display that the accretion has stopped. Radiative feedback stops accretion onto active sink particles quicker (after $\approx$ 1Myr).}
\label{fig-environment-accretion}
\end{figure}

Sink particles accrete gas from their environment as long as the gas is e.g. gravitationally bound. When accretion stops the sink particle has reached its maximum mass, $M_{\text{si, max}}$. It is useful to investigate the accretion time, $\Delta t_{\text{acc}}$, which is the time elapsed from the formation of the sink until the maximum sink mass is reached, which quantitatively demonstrates the impact of radiative feedback on the local star formation rate. In Fig. \ref{fig-environment-accretion}, we show $\Delta t_{\text{acc}}$ as a function of $M_{\text{si, max}}$ for the simulations of MC$_{1}$ (red) and MC$_{2}$ (black) with (ZI\_RAD; full markers) and without radiative feedback (ZI\_NOFB; open markers) within $V_{\text{CoV}}$. Transparent markers are sinks with masses below the massive star formation limit of 120 M$_{\odot}$. Blue crosses indicate that the accretion onto the corresponding sink has stopped. In ZI\_NOFB, the accretion times stretch over a wider temporal range and sinks grow to a few 1000~M$_{\odot}$ because accretion cannot be halted. In ZI\_RAD, the accretion time is less than $\sim$1 Myr. Sink particles with masses above the star formation threshold not only stop their own accretion but effect or even interrupt the mass accretion of any nearby companion. This results in a large fraction of sinks that remain below the star formation mass threshold. We expect that short accretion times are accompanied by a drastic change in the environmental density of the sink particles as a function of time. This is investigated in the following Section. 

\subsection{Environmental densities}
\label{sec-environment-envirdens}

Over the life time of massive stars, their environmental densities are continuously changing. The ambient density determines the impact of radiative feedback \citep{haid17}. Right after the star a star is born the surrounding gas is typically so dense, that the young \textsc{Hii} region is confined \citep{wood89,  peters10a}. The bubble then expands hydro-dynamically, while more gas is ionized. However, once the environmental density has significantly decreased (after the ionized gas has leaked out of the bubble or the star has moved out of the dense star-forming filament) the ionization front spreads out and the impact of radiative feedback is not locally confined.

\begin{figure*}
\includegraphics[width=1\textwidth]{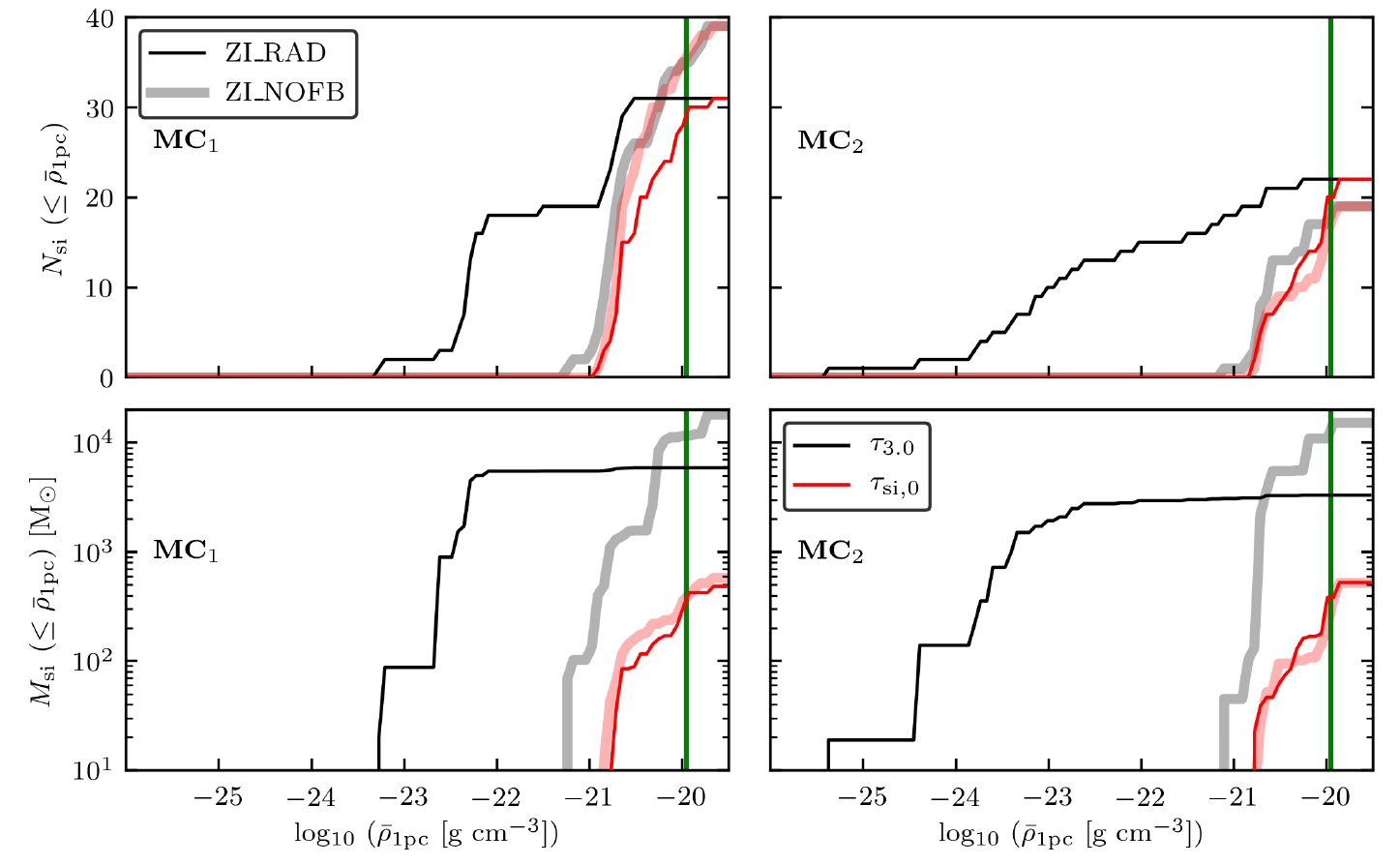}
\caption{Cumulative cluster sink particles mass $M_{\text{si}}$ (top) and number $N_{\text{si}}$ (bottom panel) distribution dependent on the environmental densities at different sink ages, $\tau_{\text{si},0}$ (red, formation time),  $\tau_{\text{si},1.0}$ (black), and $\tau_{\text{si},1.5}$ (blue) for simulation ZI\_NOFB (thick) and ZI\_RAD (thin). The top and bottom panels correspond to the total domain of MC$_{1}$ (left) and MC$_{2}$ (right), respectively. The green, vertical line shows the sink formation density threshold, $\rho_{\text{si}}$.}
\label{fig-environment-environdens}
\end{figure*}

In Fig. \ref{fig-environment-environdens}, we show the cumulative distribution of the number of sinks (top, $N_{\text{si}}$) and the cumulative sink mass (bottom, $M_{\text{si}}$) as a function of the environmental density obtained by averaging the ambient density of each sink in a sphere with a radius of 1 pc, $\bar{\rho}_{\text{1pc}}$. We show MC$_{1}$ (left) and MC$_{2}$ (right) with (ZI\_RAD; thin lines) and without feedback (ZI\_NOFB; thick lines) at two times: the formation time of a sink particle is denoted with $\tau_{\text{si,0}}$ (red), and the end time $\tau_{\rm 3.0}$ (black). The green vertical line indicates the sink formation threshold density, $\rho_{\text{si}}$.

First of all, when comparing ZI\_RAD and ZI\_NOFB in MC$_{2}$, we can see that with radiative feedback more sinks are formed, which hints towards triggered star formation. However, the higher number of sink contains a significantly lower total mass (see bottom panels), which indicates that feedback limits the accretion onto star forming dense regions (see also Fig.~\ref{fig-environment-accretion}). Furthermore, we can see that the sinks' environmental densities are severely changed by radiative feedback. All sinks are born in very dense gas ($\bar{\rho}_{\text{1pc}} >10^{-21}$ g cm$^{-3}$) and without feedback most of them also stay there (modulo some wandering off a bit). This implies that there is a large enough gas reservoir to feed the sink particles for the simulated time, even if their mass has grown significantly. 
With radiative feedback, however, the distribution is significantly shifted towards lower densities at $\tau_{\rm 3.0}$ compared to $\tau_{\text{si,0}}$. Even though there is still a number of sinks at $\bar{\rho}_{\text{1pc}} >10^{-21}$ g cm$^{-3}$, these are mostly the young sink particles which did not have time to disperse their environment. In MC$_1$ about 60 percent of all sinks are surrounded by gas with $\bar{\rho}_{\text{1pc}} < 10^{-21}$ g cm$^{-3}$. The dispersal of MC$_2$ has progressed farther and $\sim$80 percent of all sinks are found at $\bar{\rho}_{\text{1pc}} < 10^{-21}$ g cm$^{-3}$.

Fig. \ref{fig-environment-environdens} shows that the environmental densities for more than 90 and 60 percent of the sink particles in MC$_1$, respectively MC$_2$ lie between 10$^{-23}$~g~cm$^{-3}$ and 10$^{-20}$~g~cm$^{-3}$ at $\tau_{3.0}$. These are conditions, where stellar winds were shown not to be important \citep{geen15, haid17}. Therefore, we do not include this additional feedback process in this work. Nevertheless, the environmental densities are continuously reduced by ionizing radiation at later stages ($\tau > \tau_{3.0}$) stellar wind might become important. 

\subsection{The multi-phase evolution}
\label{sec-environment-evolution }
\begin{figure*}
\includegraphics[width=0.78\textwidth]{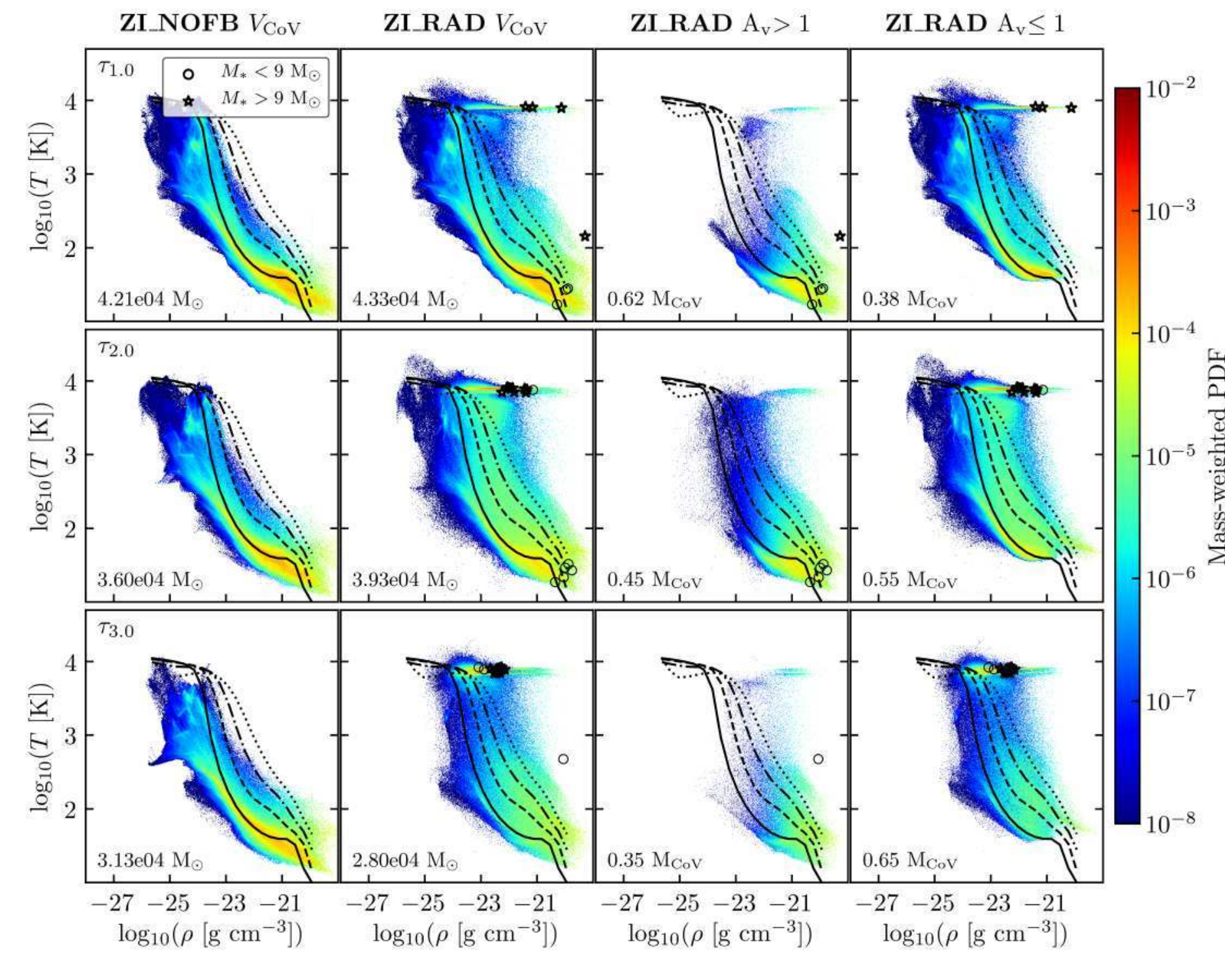}
\includegraphics[width=0.78\textwidth]{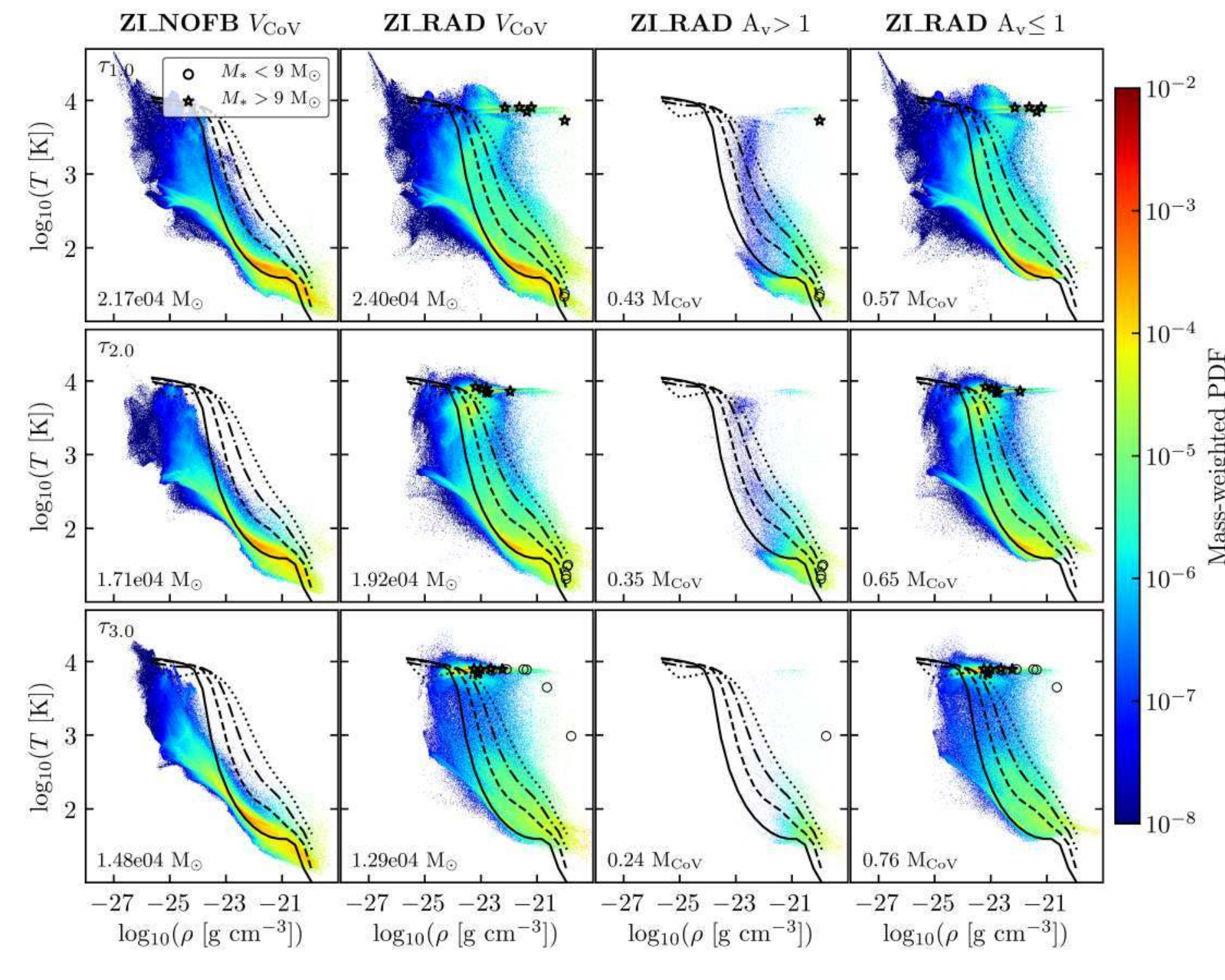}
\caption{Mass-weighted (color) density-temperature distribution of the of the central subregion ($V_{\text{CoV}}$) of MC$_{1}$ (top) and MC$_{2}$ (bottom) at times $\tau_{1.0}$ (top),  $\tau_{2.0}$ (center), and  $\tau_{3.0}$ (bottom) for simulation ZI\_NOFB (left column) and ZI\_RAD (right). The left and second column show the total gas distributions. The third and forth column show only gas above/below a an $A_{\text{v}}$ of 1 mag. The markers indicate sink particles with an active stellar component (star symbol) and without massive stars (circles). The black lines show the thermal equilibrium curve derived for G$_{0}$ of 1.7 (solid), 17 (dashed), 170 (dash-dotted), and 1700  (dotted) in units of Habing fields. The numbers in the lower left corners indicate the total gas mass in $V_{\text{CoV}}$ (first and second column) and the fraction of mass above/below $A_{\text{v}}=1$ (third and fourth column), respectively.}
\label{fig-environment-phplt-mc1}
\end{figure*}

\begin{figure*}
\includegraphics[width=0.78\textwidth]{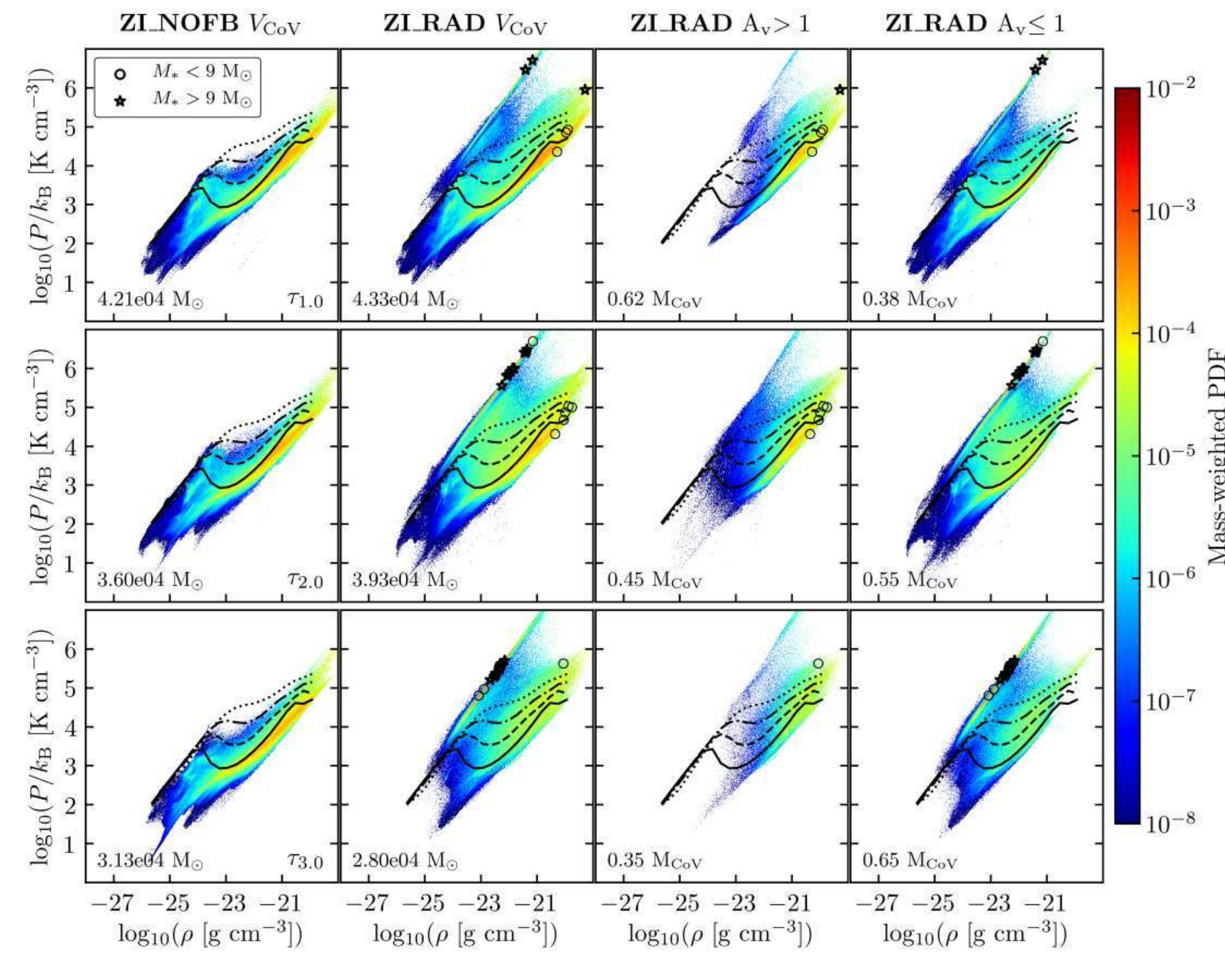}
\includegraphics[width=0.78\textwidth]{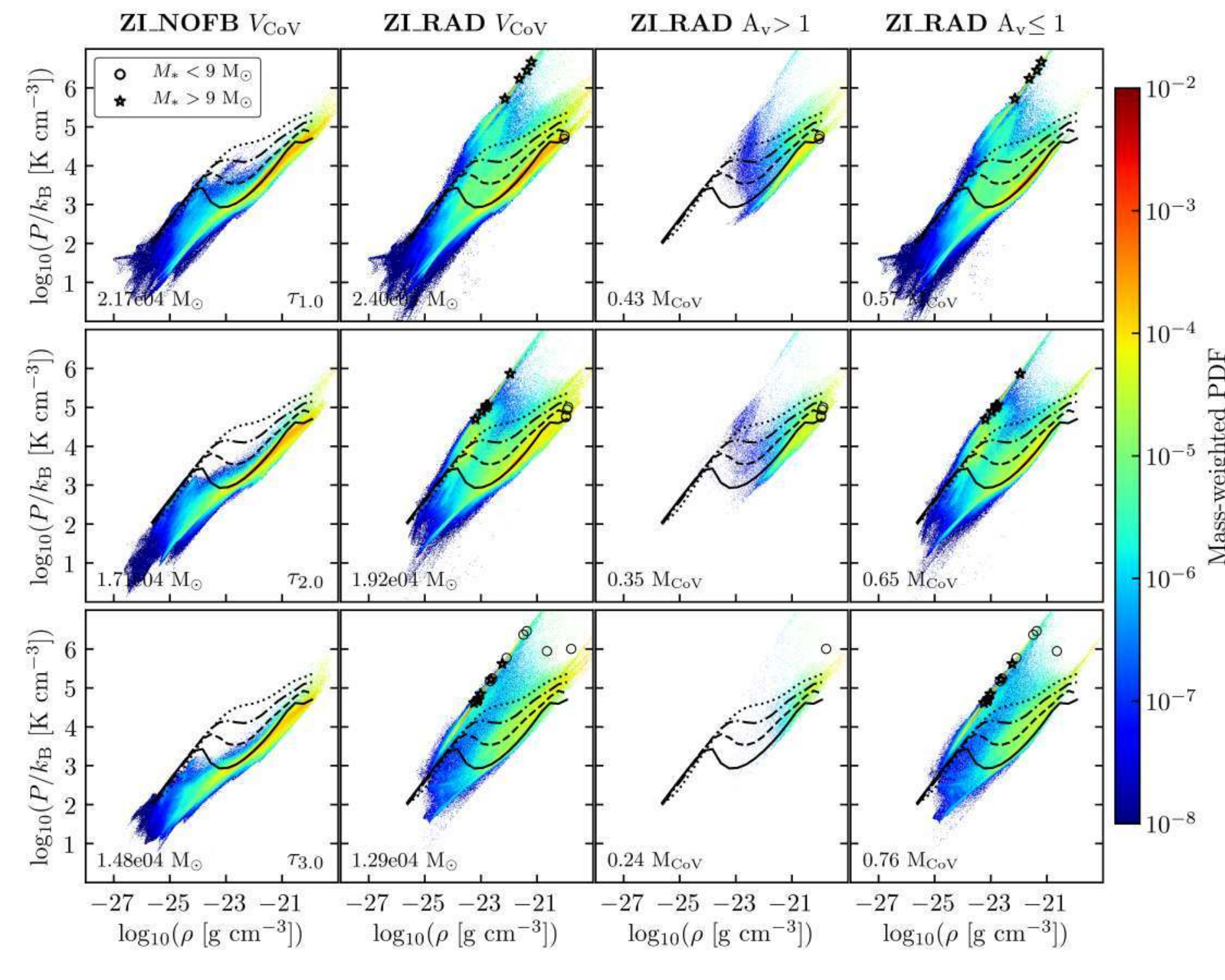}
\caption{Same figure as in Fig. \ref{fig-environment-phplt-mc1} but for the mass-weighted pressure-density distributions.}
\label{fig-environment-phplt-mc2}
\end{figure*}

We show the mass-weighted (color) density-temperature and density-pressure (pressure over the Boltzmann constant) distributions $V_{\text{CoV}}$ for both clouds in Fig. \ref{fig-environment-phplt-mc1} and Fig. \ref{fig-environment-phplt-mc2}, respectively. Note that, according to the Jeans criterion, the depicted gas density is fully resolved, even at the high density end. For each cloud (MC$_{1}$, top panels;  MC$_{2}$, bottom panels) we show different times $\tau_{1.0}$, $\tau_{2.0}$, and $\tau_{3.0}$ from top to bottom. To guide the eye, the black lines show the thermal equilibrium curves calculated using a stand-alone version of the chemistry module with increasing G$_{0}$ of 1.7 (solid), 17 (dashed), 170 (dash-dotted), and 1700 (dotted) in units of Habing fields. The thermal equilibrium curve can be assumed to be the transition to the CO dominated gas \citep{rollig07}. In each row, the left column shows the runs without radiative feedback (ZI\_NOFB) and the three panels to the right base on run ZI\_RAD. The second panel shows the total gas within $V_{\text{CoV}}$, while the third and forth panel show gas above and below a visual extinction, $A_{\text{v}} = $1 mag, which is computed self-consistently for every cell in the computational domain using the {\sc TreeRay OpticalDepth} module (see Section \ref{sec-numerical-chemistry}). The markers indicate the average environmental density, temperature, and pressure of sinks without (circles) and with active stellar components (stars) within a sphere of radius 1 pc around each sink particle, $\bar{\rho}_{\text{1pc}}$, $\bar{T}_{\text{1pc}}$, and $\overline{P/k_{\text{B}}}_{\text{1pc}}$. The numbers in the lower left corners indicate the mass within $V_{\text{CoV}}$ at given time or the fraction of mass at high and low $A_{\text{v}}$ relative to $M_{\text{CoV}}$, respectively. It can be seen that MC$_{1}$ has a significantly higher fraction of shielded gas than MC$_{2}$ (for further analysis see Section \ref{sec-extinction}). The sinks are shown in the respective high/low $A_{\text{v}}$ panels depending on their average $A_{\text{v}}$ within the surrounding 1 pc radius.

The gas distributions of runs ZI\_NOFB (Fig. \ref{fig-environment-phplt-mc1} and \ref{fig-environment-phplt-mc2}, left column) are more or less constant in time and follow the computed equilibrium curves. For $\rho \gtrsim10^{-22}$~g~cm$^{-3}$, most of the gas is more deeply embedded and cools down to $\sim$10 K. 
For runs with radiative feedback, the phase diagrams change significantly as a lot of gas is lifted above the equilibrium curve towards high temperatures and pressures. Several new horizontal branches become apparent in the temperature-density diagram. Young and deeply embedded H{\sc ii} regions first appear in the $A_{\text{v}} > $1 mag distribution (see Fig. \ref{fig-appendix-ih2ihp} for the ionization state of this gas). With time, these embedded bubbles grow, burst out of the dense filament and leak into the low density environment. Fully developed H{\sc ii} regions are heated up to temperatures between 8000~K and 10000~K occurring at $A_{\text{v}} <$ 1 mag.

Inactive stars are usually deeply embedded inside the cloud (at high density and low temperature), unless they reside near an active sink which influences their environment. The H{\sc ii} regions seem to expand until the pressure gradient between ambient medium and H{\sc ii} region across their outer boundary is diminished, which can be seen from Fig.~\ref{fig-environment-phplt-mc2} where the gas inside the H{\sc ii} region joins the rising equilibrium pressure branch of the warm ambient medium.

\section{The interaction between molecular clouds and radiative feedback}
\label{sec-difference}

MC$_{1}$ and MC$_{2}$ were chosen as two clouds with similar initial parameters \citep[see][ and Table~\ref{tab-setup1}]{seifried17}. Nevertheless, the clouds evolve differently in the presence of radiative feedback (see Fig. \ref{fig-morphology-mc1} and Fig. \ref{fig-morphology-mc2}) where MC$_{1}$ seems less affected than MC$_{2}$. In this Section, we discuss the physical property of the cloud, i.e. the local extinction, which we ultimately (after a careful and extensive analysis) identify to be responsible for the apparent differences. Next, we discuss the energy content and the star formation properties of both clouds.

\subsection{Extinction matters!}
\label{sec-extinction}

\begin{figure}
\includegraphics[width=0.5\textwidth]{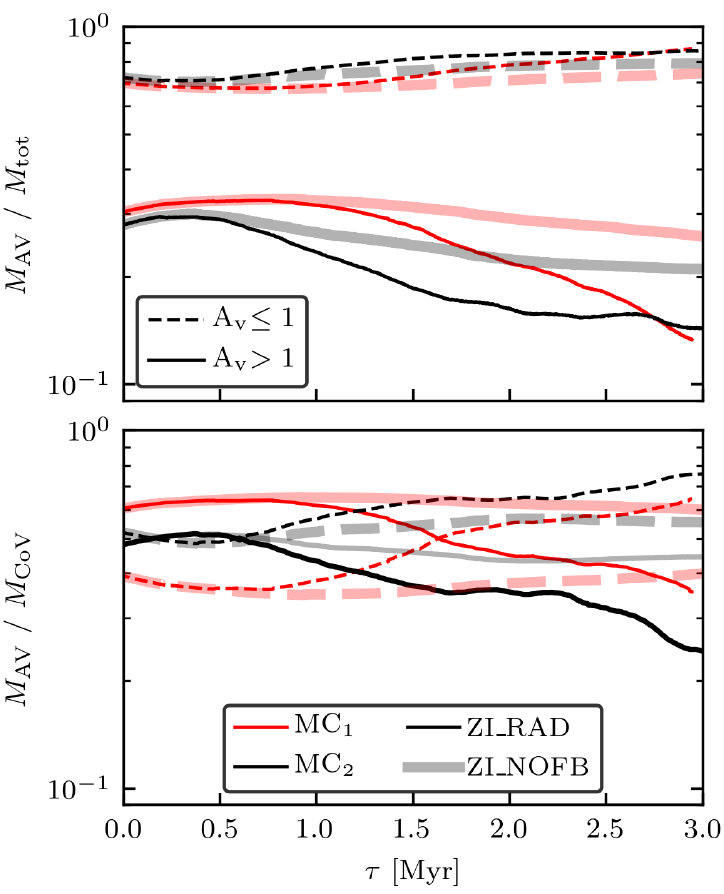}
\caption{Time evolution of the fraction of total gas mass found below (dashed) and above (solid lines) an $A_{\text{v}}$ of 1 mag for the total (top) and the central subregion (bottom) MC$_{1}$ (red) and MC$_{2}$ (black)  The differences in simulations with radiative feedback (thin lines) and without radiative feedback (thick lines) are minor, but MC$_{1}$ has more mass at $A_{\text{v}} > 1$ than MC$_{2}$. }
\label{fig-morphology-av}
\end{figure}

Fig. \ref{fig-morphology-av} shows the fraction of cloud mass constrained by different extinction thresholds in MC$_{1}$ (red) and MC$_{2}$ (black) in the total cloud ( $M_{\text{Av}}$/$M_{\text{tot}}$, top) and the central subregion ( $M_{\text{Av}}$/$M_{\text{CoV}}$, bottom) as a function of time in simulations ZI\_RAD (thin) and ZI\_NOFB (thick). Note that the evolutions of $M_{\text{tot}}$ and  $M_{\text{CoV}}$ are shown in Fig. \ref{fig-appendix-mass}. We evaluate the mass with the extinction below (dashed) and above (solid) a visual extinction of $A_{\text{v}}$ = 1~mag. For both simulations, ZI\_NOFB and ZI\_RAD, the evolutions are similar for the first 1.5 Myr. In the total domain of MC$_{1}$ and MC$_{2}$, most mass resides at $A_{\text{v}} \leq$ 1 mag with $\sim$0.7 of the total mass, M$_{\text{tot}}$. Hence, only a small fraction of the gas is well-shielded with $A_{\text{v}} >$ 1~mag and the well-shielded mass fraction is higher in MC$_{1}$ than in MC$_{2}$ by $\sim$30 percent at $\tau_{0}$ up to $\sim$80 percent at $\tau_{2.0}$ and finally become similar at $\tau_{3.0}$. In the central subregions, the evolutions of the well-shielded gas follow the larger volume with higher initial fractions of $\sim$60 percent and to $\sim$50 percent of M$_{\text{CoV}}$. During the evolution, gas is dispersed by feedback and the fraction of A$_{\text{v}}<$~1~mag dominates (compare to Fig. \ref{fig-environment-phplt-mc1} and Fig. \ref{fig-environment-phplt-mc2}).

\begin{figure*}
\includegraphics[width=0.8\textwidth]{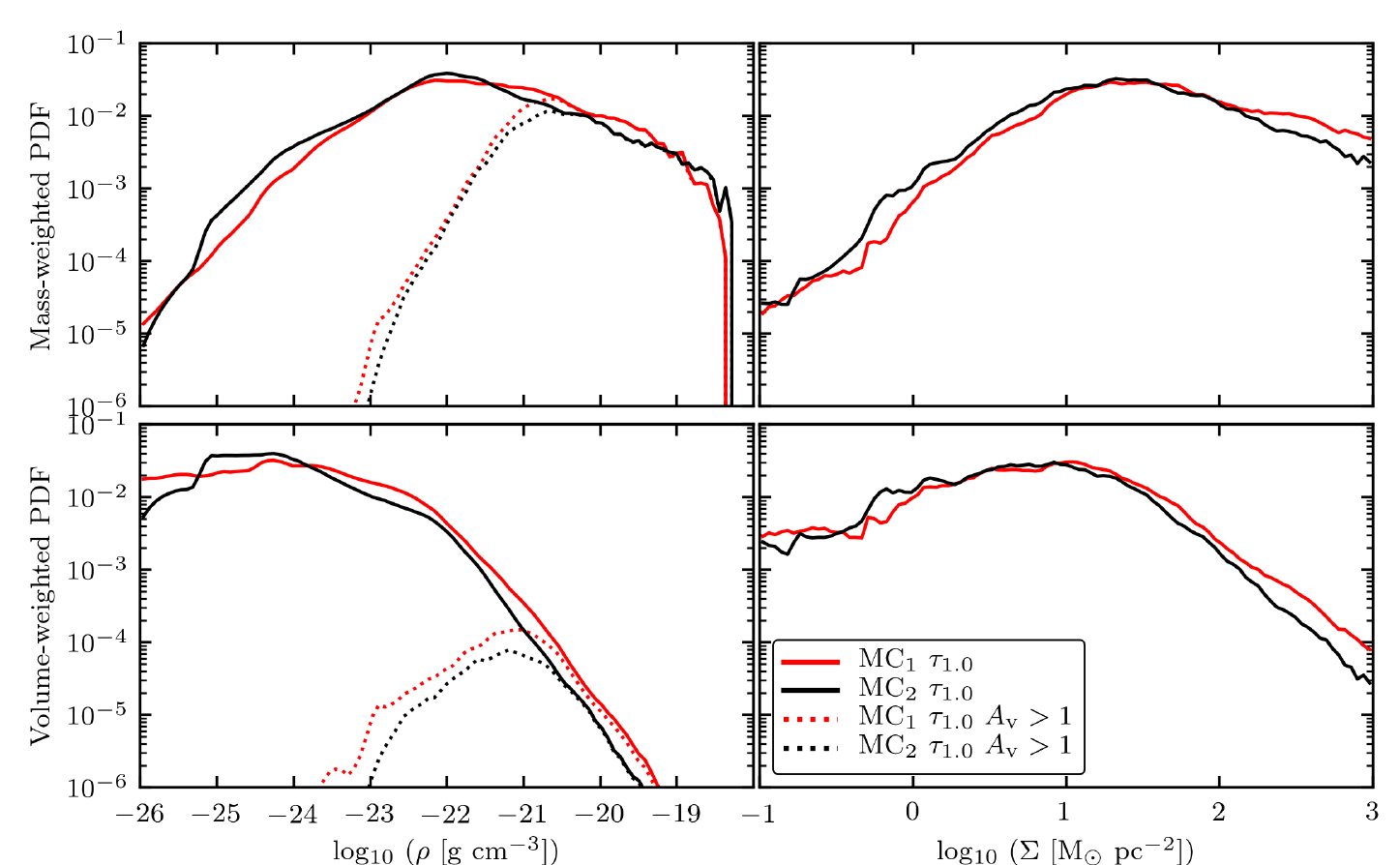}
\includegraphics[width=0.8\textwidth]{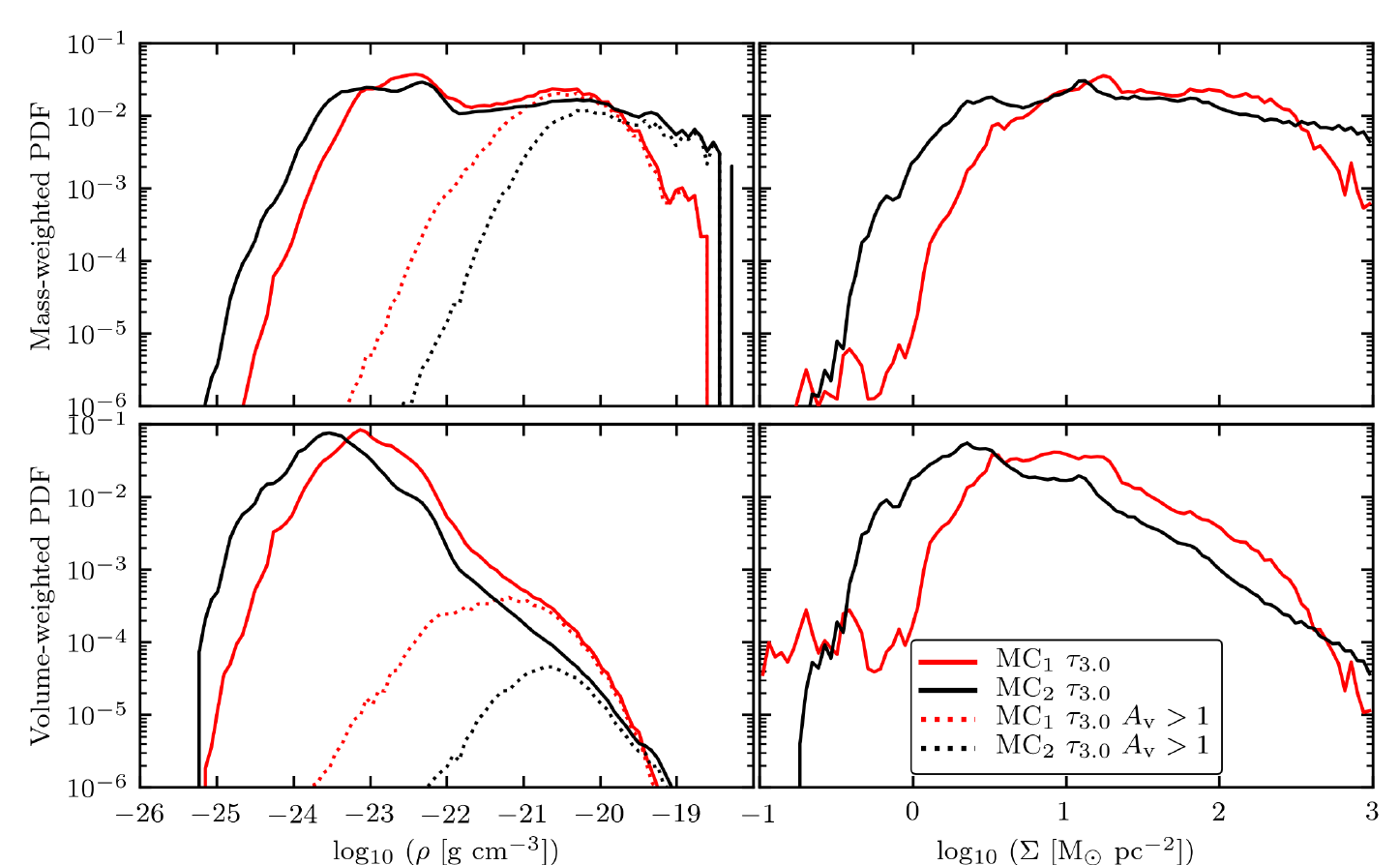}
\caption{Mass-weighted (top) and volume-weighted (bottom) density (left) and column density density PDF (right) in simulation ZI\_RAD. We show the distributions at $\tau_{1.0}$ (top four panels) and $\tau_{3.0}$ (bottom four panels) of central subregion of MC$_{1}$ (red) and MC$_{2}$ (black). Dotted lines (left column) indicate the density PDF of gas with $A_{\text{v}}>1$ mag. Although the overall PDFs of the two clouds are nearly indistinguishable, MC$_{2}$ contains a lot less well-shielded gas than MC$_{1}$. Also the dispersal of MC$_{2}$ can be seen because the fraction of gas with $A_{\text{v}}>1$ is reduced from $\tau_{1.0}$ to $\tau_{3.0}$.}
\label{fig-morphology-densityPDF}
\end{figure*}

In Fig. \ref{fig-morphology-densityPDF}, we show the mass-weighted (top) and volume-weighted (bottom) density (left) and column density PDF (right) in simulation ZI\_RAD at $\tau_{1.0}$ (top panels) and $\tau_{3.0}$ (bottom panel). The PDF includes gas within $V_{\text{CoV}}$ for MC$_{1}$ (red) and MC$_{2}$ (black). Dotted lines indicate the density PDF of gas with $A_{\text{v}}>1$~mag. This $A_{\text{v}}$ is calculated for every cell in the computational domain via the {\sc TreeRay OpticalDepth} module. We find that the mass-weighted and volume-weighted total gas density distributions of MC$_{1}$ and MC$_{2}$ are similar over a wide range of densities and at $\tau_{1.0}$ and $\tau_{3.0}$. The column density PDFs (right column) of the clouds are slightly different: While MC$_{1}$ dominates in the high $\Sigma$ regime ($\Sigma \gtrsim 100\;{\rm M}_{\odot}\;{\rm pc}^{-2}$), MC$_2$ hosts more gas at lower column densities.

The differences between the two clouds become apparent when inspecting the high-$A_{\text{v}}$ gas (blue, green). MC$_{1}$ has more mass in gas at high $A_{\text{v}}$ than MC$_2$ and this gas occupies a larger fraction of the cloud volume. Also, there is basically no difference between the two time steps $\tau_{1.0}$ and $\tau_{3.0}$ for MC$_1$, apart from the very high density tail in the mass-weighted density PDF, which forms at $\tau_{3.0}$. This indicates that the gas in MC$_1$ is still relatively confined at $\tau_{3.0}$. On the other hand, MC$_2$ has less mass at high $A_{\text{v}}$ and this mass occupies a smaller fraction of the cloud volume. Also, the mass and volume fractions of gas at high $A_{\text{v}}$ are clearly decreasing as a function of time, which is a clear sign of cloud dispersal.

Overall we find that the impact of radiative feedback is very sensitive to the detailed cloud substructure. In this regard, even if the volume density distributions are similar, the distribution of the three-dimensional, visual extinction may be different and these differences are enhanced when the cloud is exposed to radiative feedback. 

\subsection{Energy evolution}
\begin{figure}
\includegraphics[width=0.5\textwidth]{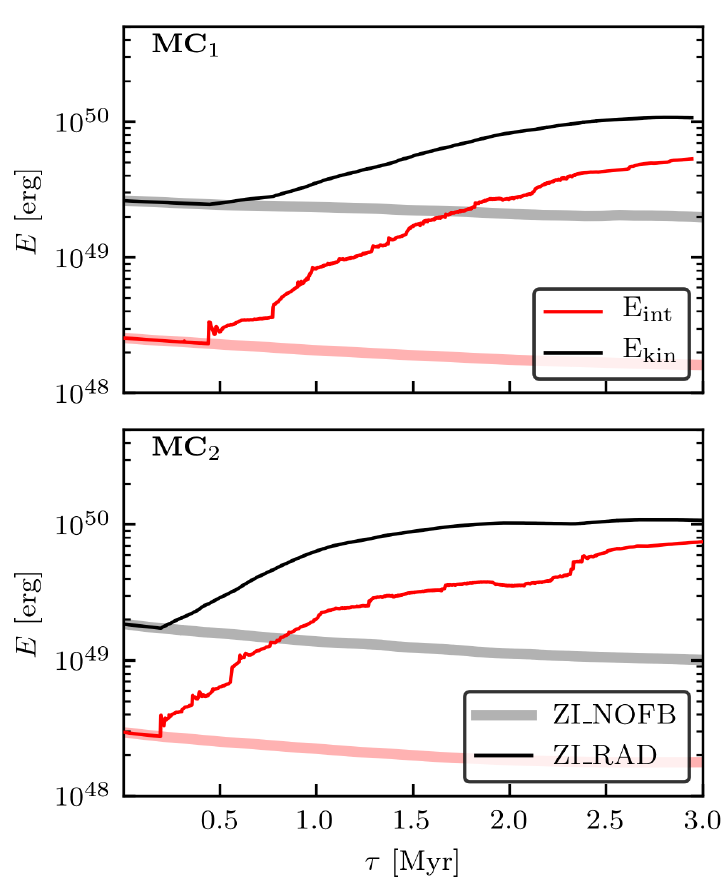}
\caption{Time evolution of the internal (red, $E_{\text{int}}$) and kinetic (black, $E_{\text{kin}}$) energy in the simulations ZI\_RAD (thin) and ZI\_NOFB (thick) in MC$_{1}$ (top) and MC$_{2}$ (bottom). The  jumps in the evolution of the internal energy are due to embedded H\textsc{ii} regions that open and release radiation into the ambient medium.}
\label{fig-morphology-ekinethegrav}
\end{figure} 

In Fig. \ref{fig-morphology-ekinethegrav}, we show the evolution of the internal (red, $E_{\text{int}}$) and kinetic (black, $E_{\text{kin}}$) energy of the gas for the simulations ZI\_RAD (thin) and ZI\_NOFB (thick) for MC$_{1}$ (top) and MC$_{2}$ (bottom). Note that we only consider the gas while the contribution to $E_{\text{kin}}$ from sink particles is neglected. The initial kinetic and internal energies are similar for both MCs with $E_{\text{int}}$ $\approx$ 3$\times$10$^{48}$ erg and $E_{\text{kin}}$ $\approx$ 2$\times$10$^{49}$ erg. Both clouds are initially bound with virial parameters of $\sim$0.72 and $\sim$0.89 as calculated for $M_{\text{tot}}$ in $V_{\text{tot}}$. In ZI\_NOFB, the energies in both clouds decrease as no feedback energy is injected. With radiative feedback the internal energy increases following the formation of massive stars. Radiative feedback also clearly enhances the kinetic energy content of the clouds, i.e. it drives turbulence \citep[see e.g.][]{gritschneder09, walch12}.

In both clouds, we see  jumps in $E_{\text{int}}$ by up to 50 percent. This behaviour is linked to the confinement of radiative bubbles. Initially, they are embedded in dense structures, only a small volume is affected and the radiative impact is delayed. But as soon as the H\textsc{ii} regions open, radiation and ionized material leak out and increase the internal energy in a larger domain. As the average rate of ionizing photons in the central volume is comparable for both clouds with a factor of $\sim$2 difference (see Fig. \ref{fig-appendix-lum}), the final energtic states are similar.

\subsection{Star formation}
\label{sec-difference-starformation}
\begin{figure}
\includegraphics[width=0.5\textwidth]{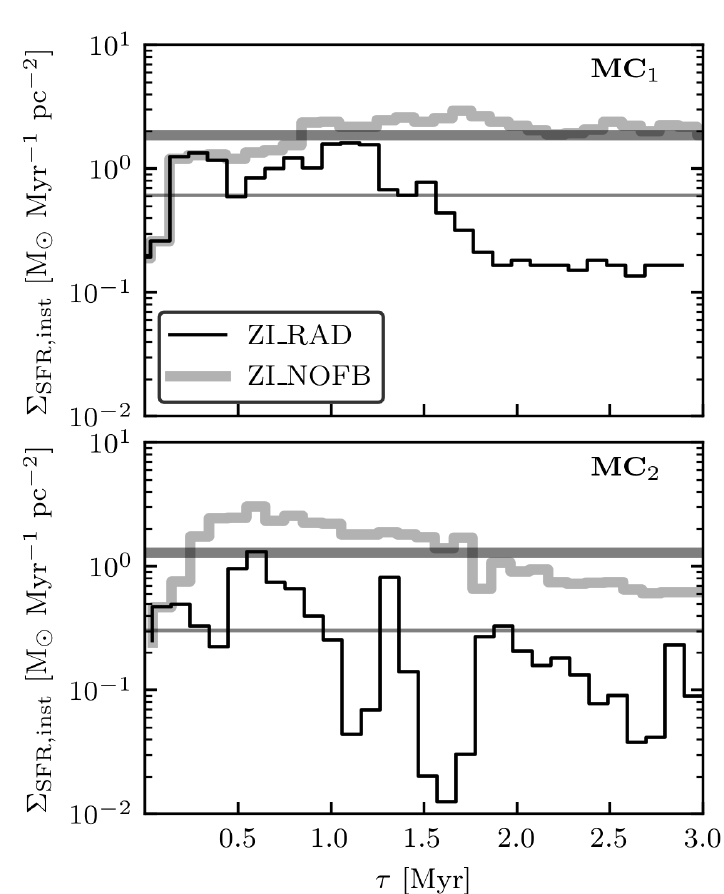}
\caption{Instantaneous star formation rate surface density $\Sigma_{\text{SFR, inst}}$ calculated for all gas within the zoom-in region of MC$_{1}$ (top) and MC$_{2}$ (bottom) that has a number density $n >$ 100~cm$^{-3}$. We show simulations with (ZI\_RAD; thin) and without radiative feedback (ZI\_NOFB; thick). Radiative feedback reduces $\Sigma_{\text{SFR, inst}}$ by a factor of $\sim$4.}
\label{fig-difference-sfrate}
\end{figure}

A common way to separate the diffuse ISM from the dense gas is to choose a density threshold. With the subscript '100' we refer to gas with number density $n >$ 100~cm$^{-3}$ ($\rho >$~3.84$\times$10$^{-22}$~g~cm$^{-3}$). To trace predominantly molecular gas, we use the subscript 'H2', which means that the mass fraction of H$_{2}$ in every cell is equal or greater than 50 percent \citep[see][]{seifried17}. A general way to indicate either of the two constraints is a subscript 'x'. 

The \textit{instantaneous star formation rate surface density} $\Sigma_{\text{SFR, inst}}$ assumes that all gas that is accreted onto a sink particle is immediately forming an ensemble of low and high mass stars. It is defined as 
\begin{equation}
\label{eq-sigma-sfr}
\Sigma_{\text{SFR, inst}} = \frac{1}{A} \sum_{j=1}^{N_{\text{si}}} \dot{M}_{si, j}(\Delta t) \left[\mathrm{M}_{\odot}\ \mathrm{Myr}^{-1}\ \mathrm{pc}^{-2} \right]
\end{equation}
where $\Delta t$ = 0.1 Myr, $A$ is the area of the cloud and $\dot{M}_{si, j}$ the mass accretion rate of the sinks over a time period $\Delta t$ \citep{matzner00, gatto17}. $A$ is calculated from the mass-weighted radius, which we calculate from the distance of all cells above a given threshold relative to the center of mass in the volume $V_{100}$.

In Fig. \ref{fig-difference-sfrate}, we depict $\Sigma_{\text{SFR, inst}}$ for MC$_{1}$ (top) and MC$_{2}$ (bottom) for the simulations ZI\_RAD (thin) and ZI\_NOFB (thick). We compute the mass-weighted cloud area from all cells with $n >$ 100~cm$^{-3}$. The time averaged values for the simulation ZI\_NOFB are 1.9 and 1.3 M$_{\odot}$ Myr$^{-1}$ pc$^{-2}$ as well as 0.6 and 0.3 M$_{\odot}$ Myr$^{-1}$ pc$^{-2}$ in run ZI\_RAD for MC$_{1}$ and MC$_{2}$, respectively. This shows that radiative feedback reduces the star formation rate surface density by a factor of $\sim$4. Similar values are obtained for a radius constrained with molecular hydrogen dominated gas. 
\begin{figure}
\includegraphics[width=0.5\textwidth]{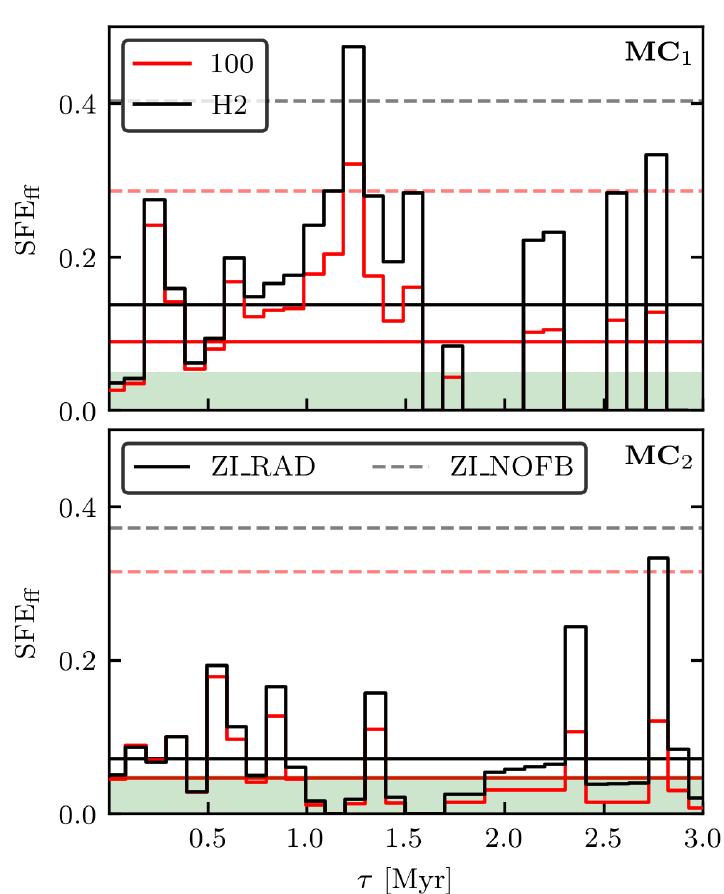}
\caption{The evolution of the star formation efficiency per free-fall time SFE$_{\text{ff,x}}$ in the total domain of MC$_{1}$ (top) and MC$_{2}$ (bottom) for the simulation ZI\_RAD with the constraints x = [100 (red), H2 (black)]. The solid, horizontal lines show the time average. The dashed horizontal lines show the time average of  the same parameter in the simulation ZI\_NOFB. The shaded green area indicates an efficiency below 5 percent.}
\label{fig-difference-stefficiency}
\end{figure} 

The \textit{star formation efficiency per free-fall time} (SFE$_{\text{ff,x}}$) is the dimensionless ratio of the mass in stars $M_{\text{st}}$ that forms within a free-fall time $\tau_{\text{ff,x}}$ divided by the mass of the cloud $M_{\text{x}}$ with x = [100, H2]. In this paper, the mass in stars is equivalent to the mass in the sink particles, $M_{\text{st}}$ = $M_{\text{si}}$. The free-fall time is given with $\tau_{\text{ff,x}} = (3\pi / (32 \text{G}\rho_{\text{x}}))^{0.5}$, where $\rho_{\text{x}} = M_{\text{x}}/V_{\text{x}}$. This gives
\begin{equation}
\label{eq-difference-epsilon}
 \text{SFE}_{\text{ff, x}}= \frac{\dot{M}_{\text{si}}}{M_{\text{x}}}\tau_{\text{ff,x}}
\end{equation}
where $\dot{M}_{\text{si}}$ = $\text{d}M_{\text{si}} / \text{d}t$ \citep{krumholz07a, murray11a, dale14}.

In Fig. \ref{fig-difference-stefficiency}, we show $\text{SFE}_{\text{ff, x}}$ for simulation ZI\_RAD in MC$_{1}$ (top) and MC$_{2}$ (bottom) for $\rho_{\text{100}}$ (red) and $\rho_{\text{H2}}$ (black). The time-averaged values are shown as horizontal, solid lines. The horizontal, dashed lines are the time-averaged values from the corresponding simulations without radiative feedback, ZI\_NOFB. The time-averaged free-fall times $t_{\text{ff,x}}$ are 2.7, 2.4 Myr in MC$_{1}$ and 2.8, 1.8 Myr in MC$_{2}$ for the thresholds x = [100, H2], respectively. For the number density threshold $\rho_{\text{100}}$ and the H$_2$-based threshold $\rho_{\text{H2}}$, the time averaged SFE$_{\text{ff}}$ are 9 and 13 percent in MC$_{1}$ and 5 and 6 percent in MC$_2$ with radiative feedback, and 29 and 40 percent in MC$_1$, respectively 31 and 37 percent in MC$_2$ without feedback. Thus, in the simulations ZI\_NOFB the average values are $\sim$4 times higher and in agreement with the findings for $\Sigma_{\text{SFR, inst}}$. In general, we obtain somewhat higher average values due to short episodes of high star formation, although the SFE$_{\text{ff,100}}$ regularly drops down to the 5-percent regime (green, shaded area). Note that SFE$_{\text{ff,H2}}$ is larger due to a smaller mass that is available for star formation (see Eq. \ref{eq-difference-epsilon}). 

Values for SFE$_{\text{ff}}$ are observed for low-mass clouds to be around a few percent \citep{krumholz07a, evans09}. In clouds which are more massive and/or have longer free-fall times, the efficiency can increase up to 30 percent \citep{murray11a}. Isolated, bound MCs in numerical simulations by \citet{dale12, dale14} show SFE$_{\text{ff}}$ without and with feedback of 16 and 11 percent, which indicates that radiative feedback is inefficient in regulating star formation. This contradicts our findings, where ionizing radiation reduces the SFE$_{\text{ff}}$ on average by a factor of 4. Similar values are found by \citet{howard16}. One reason for the inefficiency of radiative feedback in \citet{dale12} and \citet{dale14} can be found in the underlying model of \citet{diazmiller98}, which systematically underestimates the ionizing luminosities. These differ from the presented model by up to a factor of 10 lower values with increasing stellar mass (see Fig. \ref{fig-evolution-tracks}).

\begin{figure}
\includegraphics[width=0.5\textwidth]{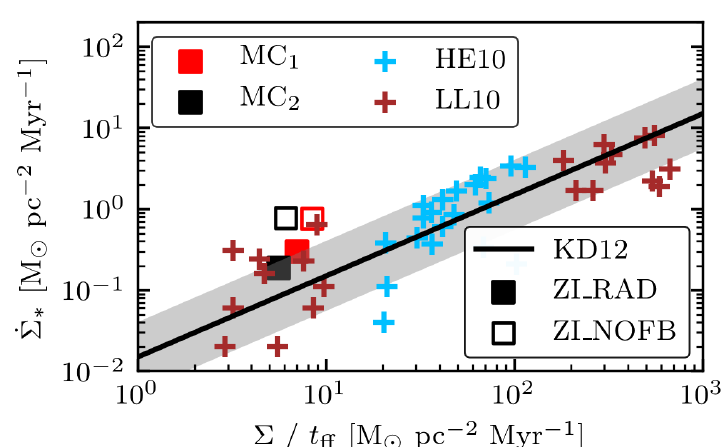}
\caption{Relation between the star formation surface density $\dot{\Sigma}_{\ast}$ and surface density over the free-fall time $\Sigma/t_{\text{ff}}$ for MC$_{1}$ (red square) and MC$_{2}$ (black square) for the simulations ZI\_RAD (full markers) and ZI\_NOFB (empty markers) at $\tau_{3.0}$. The black line shows the fitted behaviour found in \citet{krumholz12} surrounded by the scatter in grey. The data is taken from \citet{ll10} and \citet{he10}. Our simulated clouds are closest to three nearby low-mass star forming molecular clouds, Taurus, Ophiuchus, and Lupus 3. }
\label{fig-starformation-krumholz}
\end{figure}

Fig. \ref{fig-starformation-krumholz} compares the SFE$_{\text{ff}}$ obtained from the simulations with resolved observations of Milky Way molecular clouds, kpc-scale observations of Local Group galaxies, and from unresolved observations of both disk and starburst galaxies in the local universe and at high redshift published in \citet{ll10} and \citet{he10}. We relate the star formation rate surface density $\dot{\Sigma}_{\ast} = \Sigma_{\text{SFR, inst}}(\Delta t = 3\_\text{Myr})$ with the surface density over the free-fall time $\Sigma/t_{\text{ff}}$ derived for MC$_{1}$ (red) and MC$_{2}$ (black) for the simulations ZI\_RAD (full markers) and ZI\_NOFB (empty markers) at $\tau_{3.0}$. For each cloud, we only consider gas within its V$_{\rm tot}$ above the number density threshold $n >$ 100~cm$^{-3}$. The black line and grey shaded area shows the fitted behaviour found in \citet{krumholz12, Krumholz2013} and the associated uncertainty. 

The SFE$_{\text{ff}}$ obtained for the ZI\_NOFB runs are too high and clearly offset from the observed relation. However, both clouds with radiative feedback are right on top of the relation, with MC$_2$ at slightly lower $\dot{\Sigma}_{\ast}$ than MC$_1$. The two observed points which sit directly on top of our results correspond to the Taurus and Ophiuchus MCs (points near MC$_2$) and Lupus 3 (point on top of MC$_1$ result). This result is reassuring because these observed clouds lie in the solar neighbourhood - the environment simulated here - and have total masses and other physical properties that are comparable to our simulated clouds.

\section{Discussion: Differences between MC$_{1}$ and MC$_{2}$}
\label{sec-difference-mc1mc2}

MC$_{1}$ and MC$_{2}$ condense out of the same multi-phase ISM. They were selected to have similar initial properties like masses around 10$^{4}$ M$_{\odot}$, similar volumes, comparable kinetic and internal energies, and similar virial parameters of 0.72 and 0.89. However, during the evolution under the influence of ionizing radiation both clouds seem to diverge in respect to their morphologies, meaning that most of the gas in MC$_{1}$ remains (see Fig. \ref{fig-morphology-mc1}) while MC$_{2}$ is almost fully dispersed at $\tau_{\rm 3.0}$ (see Fig. \ref{fig-morphology-mc2}). 

In Fig. \ref{fig-environment-environdens}, we show that the environmental densities of sink particles gradually decrease with age and particularly the sinks in MC$_{2}$ are embedded in low-density media. Together with Fig. \ref{fig-morphology-densityPDF}, where we show that this cloud has much less well-shielded gas with $A_{\text{v}}>$~1~mag, we interpret that MC$_{1}$ has more deeply embedded dense structures and a thicker envelope. The density-temperature distribution of the central region of the clouds (see Fig. \ref{fig-environment-phplt-mc1}) with $A_{\text{v}}>$~1~mag indicates that some sources are deeply embedded in these well-shielded regions. Therefore, radiative feedback is confined to small bubbles in MC$_1$. The radiative impact is delayed until the radiative bubbles open into the ambient medium, ionized gas and radiation leak out and induce kinetic motions.

It is important to mention, that the emitted radiative energy is similar in both central regions (see Fig. \ref{fig-appendix-lum} for the rate of UV-photons). There are also massive stars forming in the rest of the cloud. The most relevant stars have high masses. In our simulations, almost all of those are situated far away from the central subregion. With distance and with decreasing mass the UV-photon rates per volume decrease and easily drop below the rates form a 9 M$_{\odot}$, hence are considered as minor. The effect of feedback from massive stars outside the subregions on the dense structures is minor. Otherwise it should also be visible in the mass evolution of well shielded gas (see Fig. \ref{fig-morphology-av} and compare ZI\_NOFB with ZI\_RAD) and in the density-temperature distribution, but both remain almost unchanged (see Fig \ref{fig-environment-phplt-mc1} and Fig. \ref{fig-environment-phplt-mc2}). 

Concerning the star formation in both clouds, ionizing radiation is able to lower the star formation rate surface density by a factor of $\sim$4. The star formation efficiency constrained by gas above 100~cm$^{-3}$ is found to be on average $\sim$5~--~9 percent in both clouds (see Fig. \ref{fig-difference-stefficiency} and Fig. \ref{fig-difference-sfrate}). In ZI\_NOFB, a few, massive sink particles evolve, whereas in ZI\_RAD in cloud MC$_{2}$ the sink masses are significantly reduced but their number increased.  Star formation is triggered by radiative feedback. 

The comparison of MC$_{1}$ and MC$_{2}$ shows that, not only the cloud masses \citep{dale12, dale13}, the corresponding luminosities \citep{geen18} and escape velocities influence the impact from radiative feedback, but that the initial cloud sub-structure significantly determines the cloud evolution. The initial conditions are imprinted during the formation process of the cloud \citep{brunt09, reyraposo17}. We find that the fully three-dimensional shielding properties determine the time scales of molecule formation \citep{seifried17} as well as the time scales for cloud dispersal (this paper).


\section{Conclusions}
\label{sec-summary}

In this paper, we investigate the impact of ionizing radiation feedback from massive stars in the early evolution of MCs up to 3 Myr. We perform hydrodynamic simulations with the AMR code \textsc{FLASH} 4 and include the novel radiative transfer scheme \textsc{TreeRay}, which is coupled to a chemical network to treat the effect of ionizing radiation. We self-consistently follow the formation of two, initially bound MCs from a SN-driven, multi-phase ISM down to a resolution of 0.122 pc within the SILCC-ZOOM project. We allow for sink particle formation on the highest refinement level. In the simulation ZI\_RAD, ionizing radiation is coupled to massive stars. Simulation ZI\_NOFB is the reference run without feedback process. In the following we list the main conclusions.
\begin{itemize}
\item Despite the similar initial masses of the two MCs, the morphological evolutions under the influence of ionizing radiations is different. In MC$_{1}$, a central blob of gas remains, whereas a part of MC$_{2}$ is fully dispersed. We show that this difference is linked to the mass of internal substructures of dense and well shielded gas, which embeds, and delays radiative feedback. The substructures are imprinted during the formation of the clouds. 
\item We show that the total gas density PDFs are nearly identical for the different MCs. However, the well-shielded gas ($A_{\text{v}}  >$~1~mag) reveals cloud-specific properties. In this work, MC$_{1}$ shows more volume-filling gas at intermediate densities, i.e. a thicker envelope surrounding the densest, star-forming filaments. These are responsible for sustaining the cloud structure despite the impact of radiative feedback. MC$_{2}$ contains less of this gas, hence becomes more easily dispersed. 
\item In the simulation, we find for some massive stars that the environmental densities are high and the radiative bubbles are embedded in substructures. When the radiative bubble opens into the ambient medium, internal energy and hot, ionized gas is released and the embedded phase is terminated. This behaviour is reflected by small jumps in the internal energy evolution. In this phase the ionized gas inside the H\textsc{ii} region is heated to the prototypical $\sim$8000 K. 
\item Star formation can be regulated by radiative feedback. In simulations ZI\_RAD the star formation efficiency drops by a factor of $\sim$4 compared to the ZI\_NOFB in both clouds. The star formation efficiency in gas with densities above 100 cm$^{-3}$ lies at $\sim$5~--~9 percent. This indicates that internal morphologies regulate the impact of photo-ionizing radiation, hence the star formation. 
\item Without feedback, a few sink particles accrete a significant fraction of the cloud mass. Radiative feedback significantly reduces the sink mass and instead may increase its number. This seems to be triggered star formation, even though the overall star formation efficiency is so severely reduced. 
\item When comparing with observational data, we find that our two clouds, which were simulated using typical solar neighbourhood conditions, lie on top of the results derived for Taurus, Ophiuchus, and Lupus 3, three low-mass, star forming, nearby molecular clouds with similar total masses.
\end{itemize}

\section{Acknowledgements}

SH, SW, DS and FD acknowledge the support by the Bonn-Cologne Graduate School for physics and astronomy which is funded through the German Excellence Initiative.  SH, SW, and DS also acknowledge funding by the Deutsche Forschungsgemeinschaft (DFG) via the Sonderforschungsbereich SFB 956 ''Conditions and Impact of Star Formation" (subproject C5). SH, SW, DS, FD and TN acknowledge the support by the DFG Priority Program 1573 ''The physics of the interstellar medium". SH and SW acknowledge funding by the European Research Council through ERC Starting Grant No. 679852 ''RADFEEDBACK". TN acknowledges support from the DFG cluster of excellence ''Origin and Structure of the Universe". R.W. acknowledges support by the Albert Einstein Centre for Gravitation and Astrophysics via the Czech Science Foundation grant 14-37086G and by the institutional project RVO:67985815 of the Academy of Sciences of the Czech Republic. The software used in this work was developed in part by the DOE NNSA ASC- and DOE Office Science ASCR-supported \textsc{FLASH} Center for Computational Science at University of Chicago. The authors gratefully acknowledge the Gauss Centre for Supercomputing e.V. (www.gauss-centre.eu) for funding this project (pr62su) by providing computing time on the GCS Supercomputer SuperMUC at Leibniz Supercomputing Centre (www.lrz.de). We thank the \textsc{YT-PROJECT} community \citep{turk11} for the \textsc{YT} analysis package, which we used to analyse and plot most of the data. We thank the anonymous referee for the constructive input.

\bibliographystyle{aa}
\bibliography{GeneralBib}
\newpage

\appendix

\section{Mass evolution}
\begin{figure}
\includegraphics[width=0.5\textwidth]{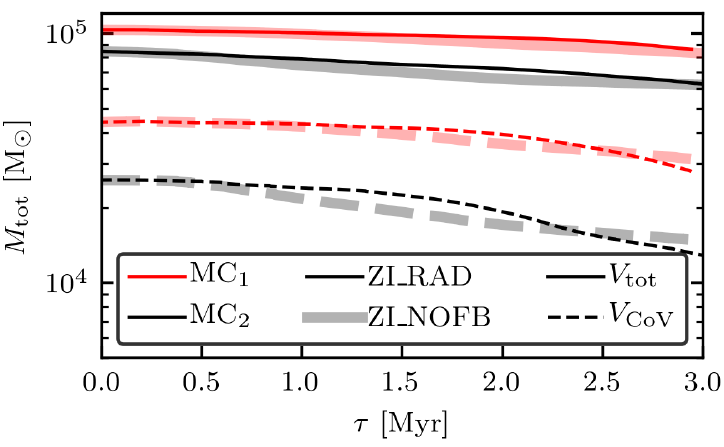}
\caption{Evolution of the mass, $M_{\text{tot}}$, in the total volume, $V_{\text{tot}}$ (solid), and the central subregion, $V_{\text{CoV}}$ (dashed), in MC$_{1}$ (red) and MC$_{2}$ (black) for simulation ZI\_RAD (thin) and ZI\_RAD (thick). }
\label{fig-appendix-mass}
\end{figure}

Fig. \ref{fig-appendix-mass} shows the time evolution of the total mass in MC$_{1}$ (red) and MC$_{2}$ (black) for simulation ZI\_RAD (thin) and ZI\_NOFB (thick) in the total volume, $V_{\text{tot}}$ (solid), and the central subregion, $V_{\text{CoV}}$ (dashed).

\begin{figure}
\includegraphics[width=0.5\textwidth]{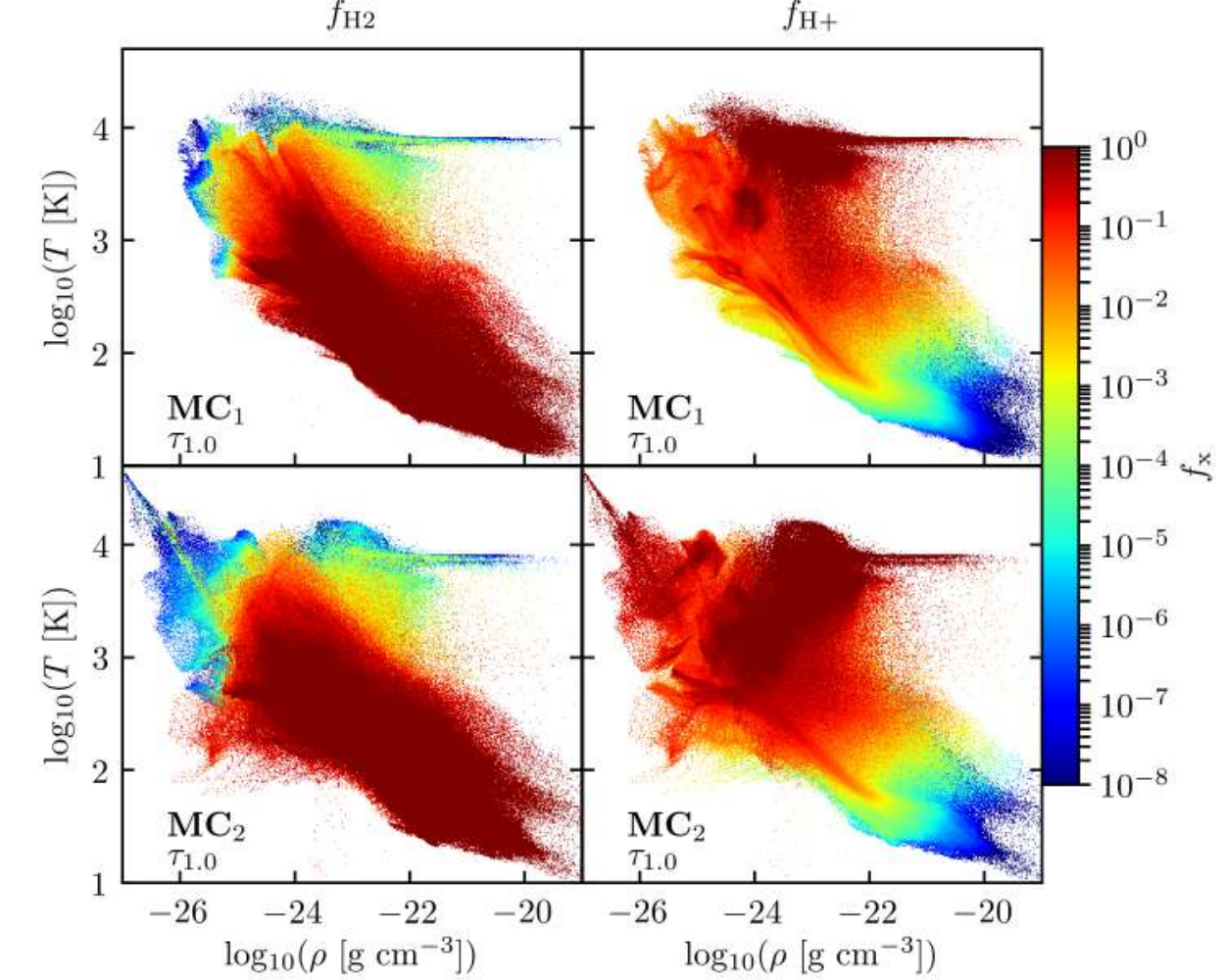}
\includegraphics[width=0.5\textwidth]{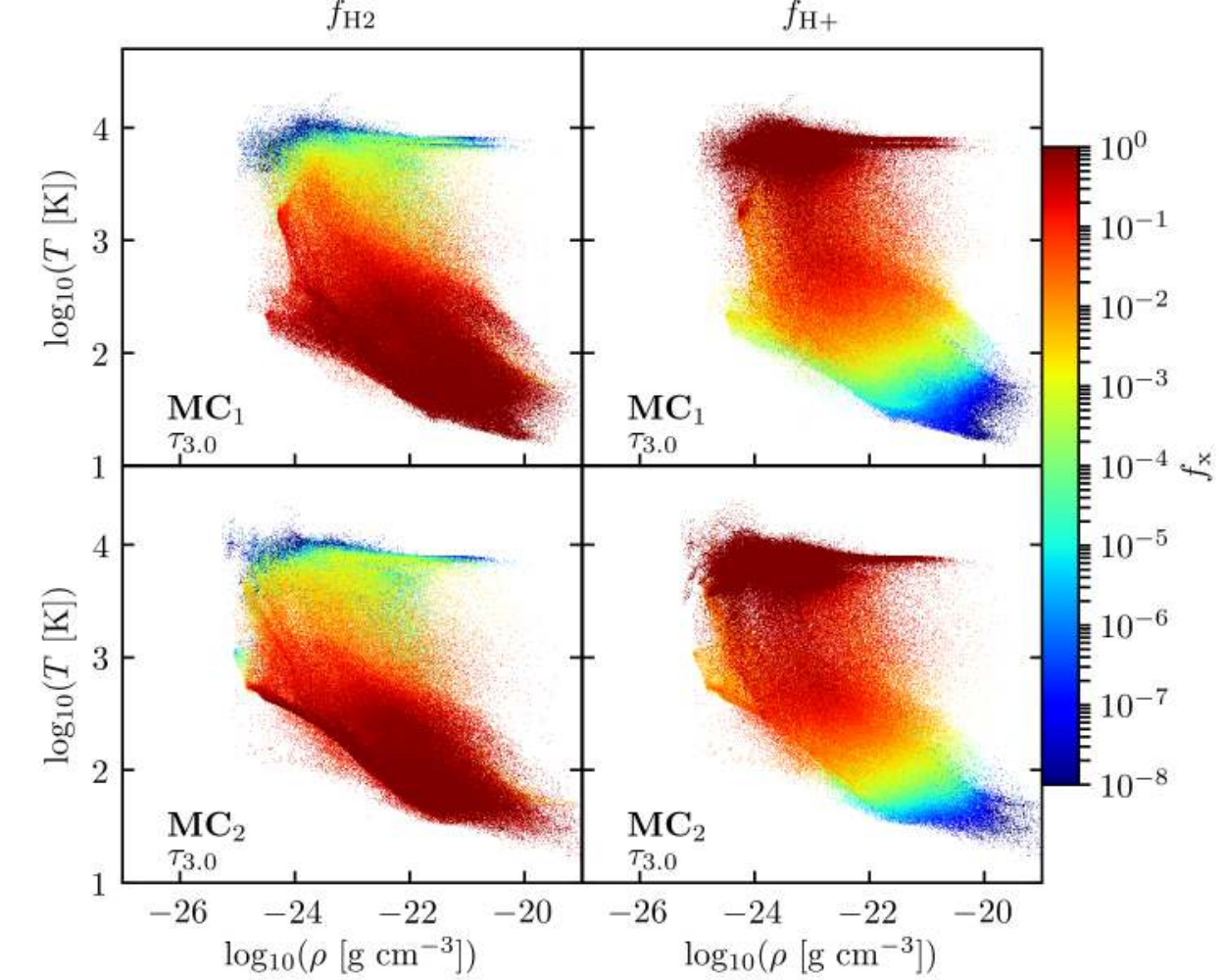}
\caption{Density-temperature distribution of MC$_{1}$ (top) and MC$_{2}$ (bottom) with the molecular hydrogen fraction (left) and ionized hydrogen fraction (right) at the $\tau_{1.0}$ (top panels) and $\tau_{3.0}$ (bottom panels).}
\label{fig-appendix-ih2ihp}
\end{figure}

Fig. \ref{fig-appendix-ih2ihp} shows the density-temperature distribution of MC$_{1}$ (top subpanel) and MC$_{2}$ (bottom subpanel) at $\tau_{1.0}$ (top) and $\tau_{3.0}$ (bottom). The molecular hydrogen fraction (left) and ionized hydrogen fraction (right) is indicated by color. 

\section{Stellar mass evolution}
\begin{figure}
\includegraphics[width=0.5\textwidth]{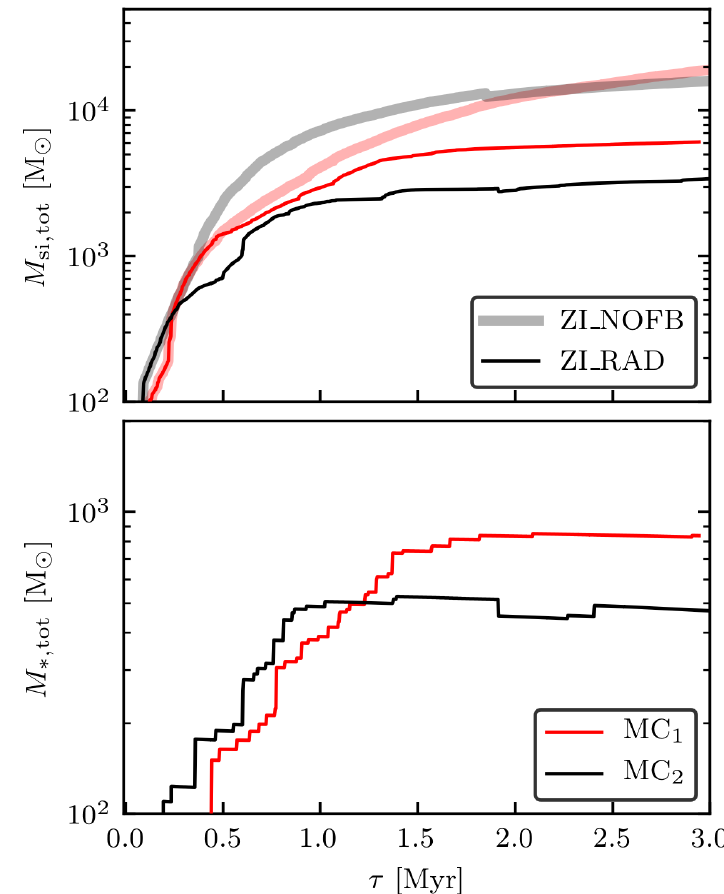}
\caption{Evolution of the sink mass $M_{\text{si, tot}}$ (top) and the stellar mass $M_{\ast,\text{tot}}$ (bottom) in MC$_{1}$ (red) and MC$_{2}$ (black) for simulation ZI\_RAD (thin, only top) and ZI\_RAD (thick). The evolution of $M_{\ast,\text{tot}}$ shows some oscillation at the beginning, which is due to stars, that move out of the molecular cloud, hence are not considered in the analysis.}
\label{fig-appendix-sistmass}
\end{figure}

Fig. \ref{fig-appendix-sistmass} shows the time evolution of the sink mass, $M_{\text{si, tot}}$, (top) and the total mass of massive stars, $M_{\ast,\text{tot}}$, (bottom) in the total volume $V_{\text{tot}}$ of  MC$_{1}$ (red) and MC$_{2}$ (black) for simulation ZI\_NOFB (thick, only top) and ZI\_RAD (thin). The initial phase of $M_{\ast,\text{tot}}$ shows oscillations, which is due to two particles, which move out of the domain.
\begin{figure}
\includegraphics[width=0.5\textwidth]{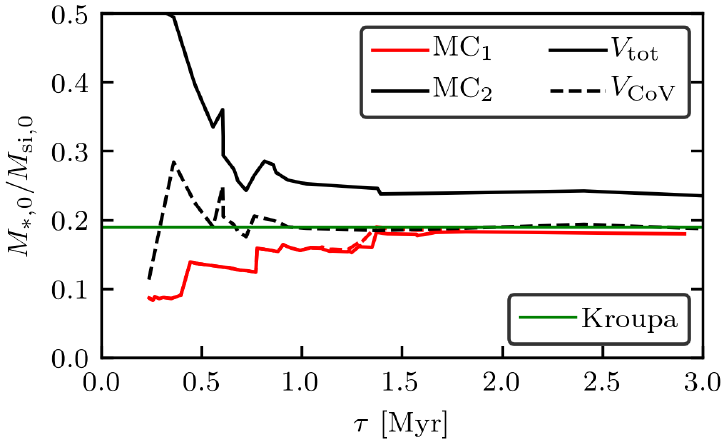}
\caption{Evolution of the ratio between the stellar and sink mass at the time of stellar formation in MC$_{1}$ (red) and MC$_{2}$ (black) for simulation ZI\_RAD in the total cloud (solid) and the central subregions (dashed). The mass fraction of the high mass range in respect to the underlying Kroupa IMF is shown as a horizontal, green line.  Note that after about 2 Myr star formation has ceased and the mass fraction remains therefore constant.}
\label{fig-appendix-revision1}
\end{figure}

\begin{figure}
\includegraphics[width=0.5\textwidth]{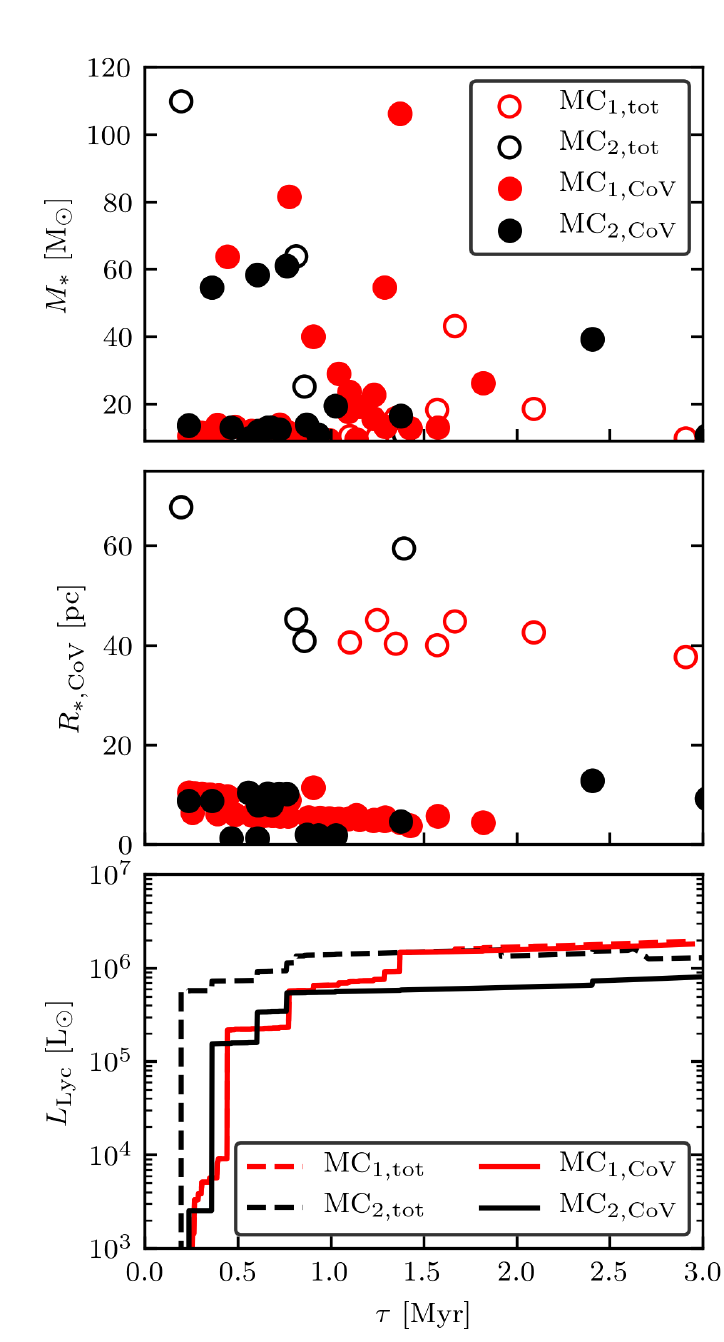}
\caption{The mass of new-born massive stars at their birth time (top), and their distance to the center of mass (middle panel) of the clouds MC$_{1}$ (red) and MC$_{2}$ (black) as well as the time evolution of the luminosity of Lyman continuum photons (bottom). Full markers and solid lines indicate that the star is located in the central volume with (40 pc)$^3$, whereas empty markers and dashed lines indicate indicate a position within the rest of the cloud.}
\label{fig-appendix-lum}
\end{figure}

Fig. \ref{fig-appendix-lum} shows the time evolution of the initial stellar mass (top), the distance (center) to the center of the small, central subregion of cloud MC$_{1}$ (red) and MC$_{2}$ (black) and the luminosity of Lyman continuum photons (bottom). Full markers and solid lines and empty markers and dashed lines indicate that the star is located in the central subregion or the rest of the cloud, respectively.

Fig. \ref{fig-appendix-revision1} shows the ratio between the stellar and sink mass at the time of the stellar formation, $M_{\mathrm{\ast, 0}}$ / $M_{\mathrm{si, 0}}$, in MC$_{1}$ (red) and MC$_{2}$ (black) for simulation ZI\_RAD in the total cloud (solid) and the central subregions (dashed). The mass fractions of the high mass range in respect to the underlying Kroupa IMF lies at 18 percent (green horizontal line). Hence, a stellar population (in a cloud) which satisfies this ratio represents the IMF well. Ratios above and below indicate that massive stars over-, respectively under-represented the IMF. After an initial massive star deficit, both central subregions and MC2 are well sampling the IMF. The high values in MC1 are caused by an initial $\approx$ 100 M$_{\odot}$ star.

\section{Other Projections}
\begin{figure}
\includegraphics[width=0.5\textwidth]{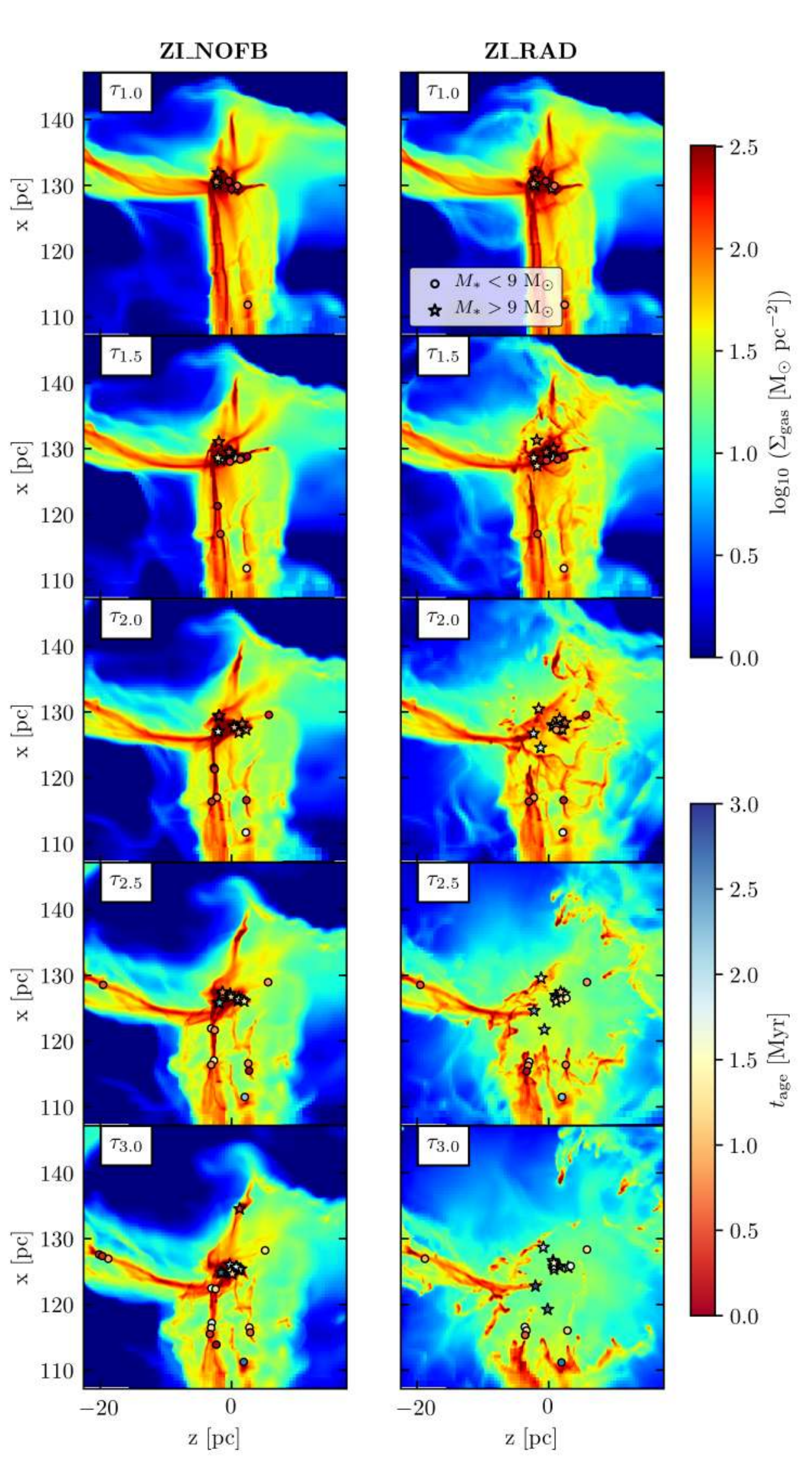}
\caption{Same as the left two columns in Fig. \ref{fig-morphology-mc1} but projected along y-direction.}
\label{fig-appendix-proj-mc1y}
\end{figure}

\begin{figure}
\includegraphics[width=0.5\textwidth]{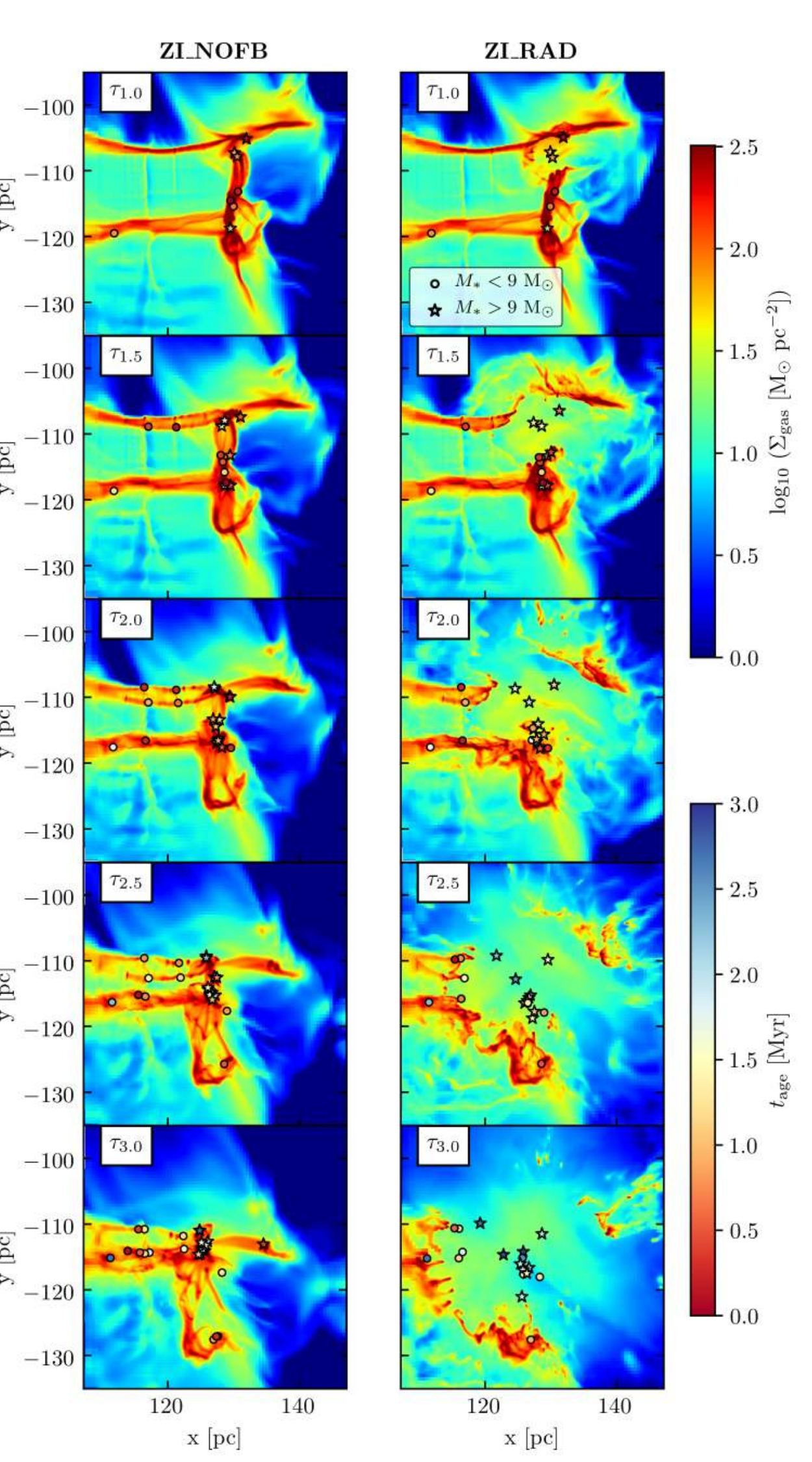}
\caption{Same as the left two columns in Fig. \ref{fig-morphology-mc1} but projected along z-direction.}
\label{fig-appendix-proj-mc1z}
\end{figure}

\begin{figure}
\includegraphics[width=0.5\textwidth]{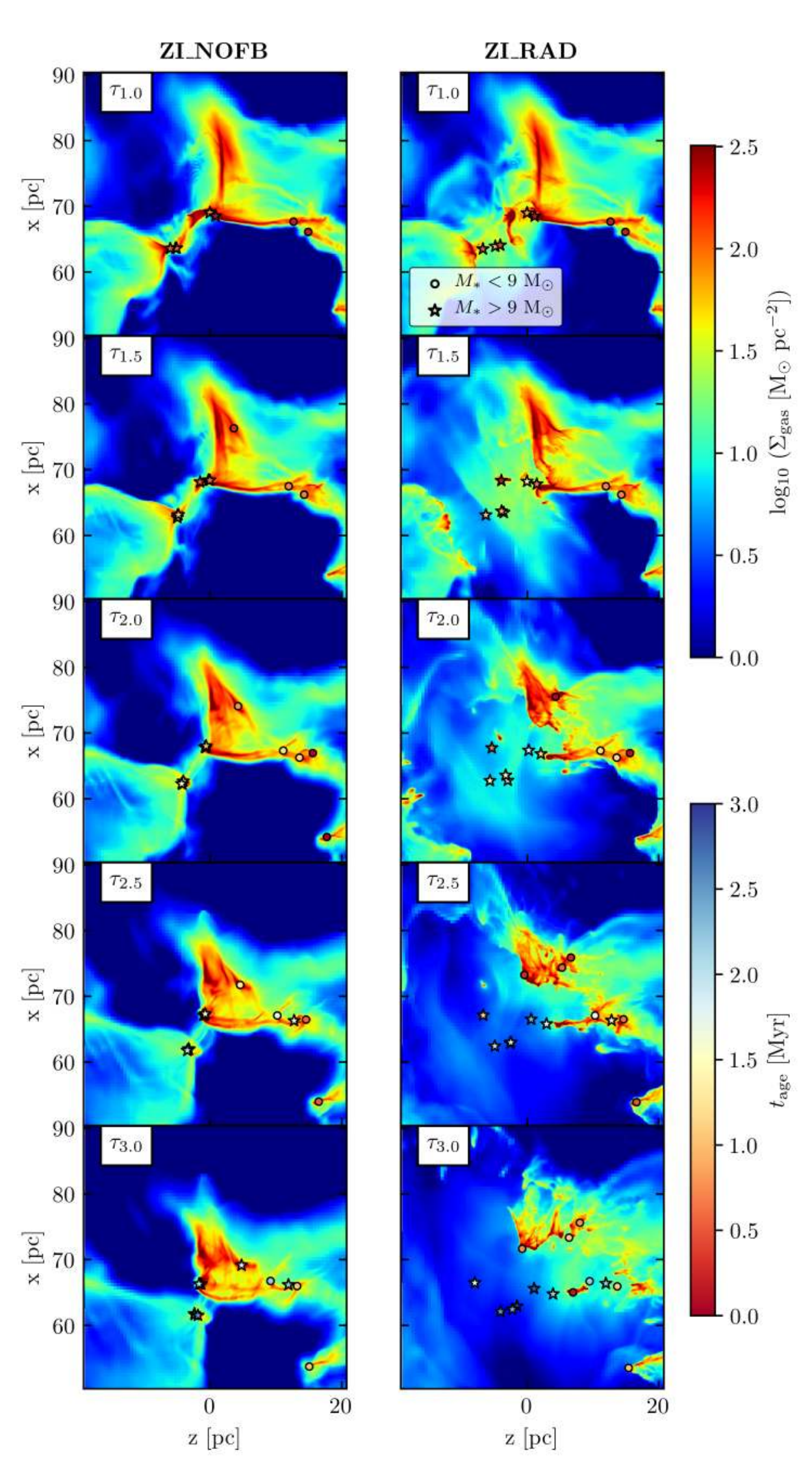}
\caption{Same as the left two columns in Fig. \ref{fig-morphology-mc2} but projected along y-direction.}
\label{fig-appendix-proj-mc2y}
\end{figure}

Fig. \ref{fig-appendix-proj-mc1y} and  \ref{fig-appendix-proj-mc1z} as well as Fig. \ref{fig-appendix-proj-mc2y}, \ref{fig-appendix-proj-mc1z} show the time evolution of the gas column density $\Sigma_{\text{gas}}$ in the $x$-$z$-plane and in the $x$-$y$ plane for the central subregion ($V_{\text{CoV}}$, see Fig. \ref{fig-setup-init} and Table \ref{tab-setup1}) of MC$_{1}$ and MC$_{2}$,respectively. The two columns show simulations ZI\_NOFB (left) and ZI\_RAD (right), respectively, at times $\tau_{1.0}$ to $\tau_{3.0}$ (from top to bottom). Circles indicate sink particles without an active stellar component, i.e. without massive stars. Star-shaped markers are cluster sink particles with active stellar feedback. The color of the markers indicate the sink age. 

\begin{figure}
\includegraphics[width=0.5\textwidth]{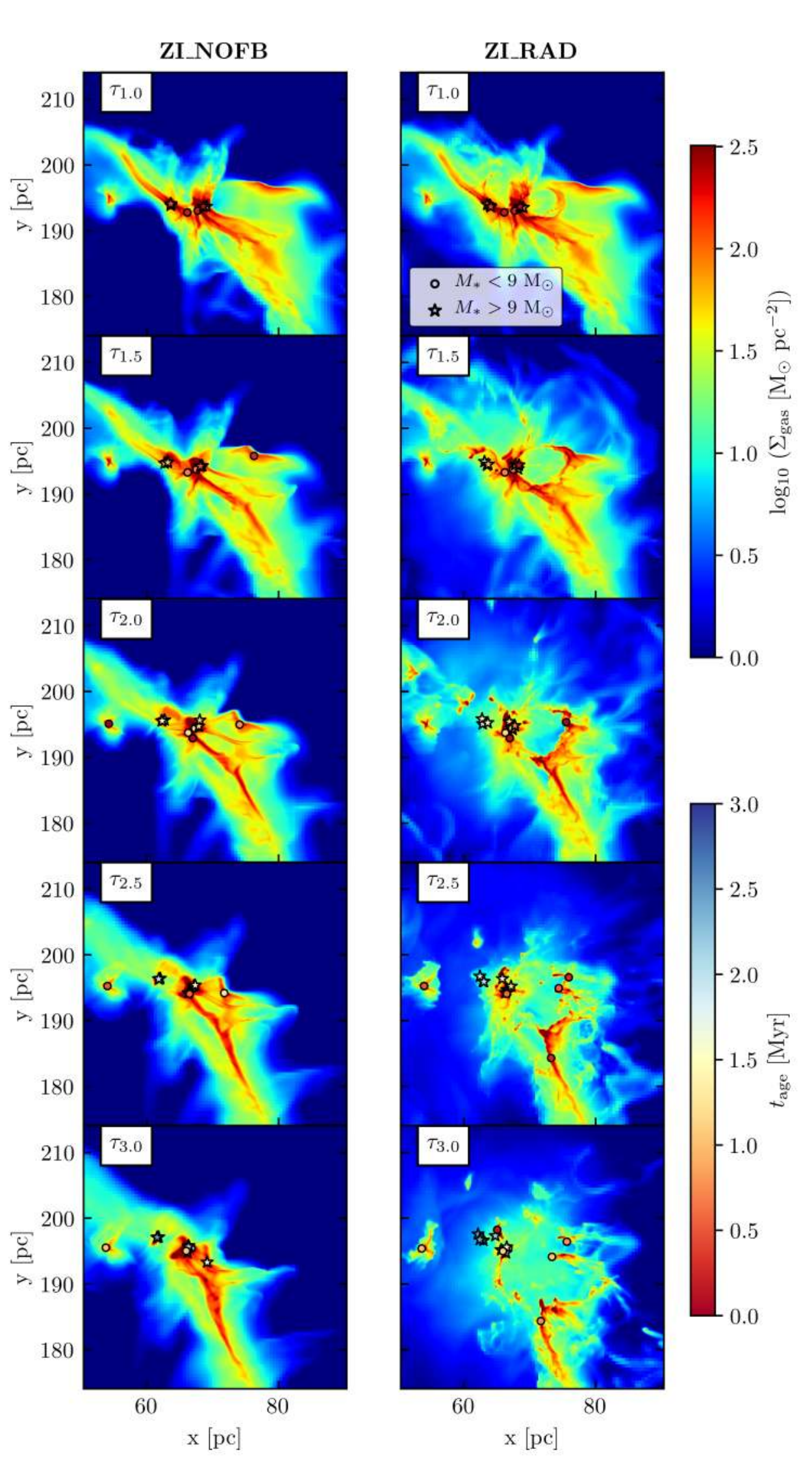}
\caption{Same as the left two columns in Fig. \ref{fig-morphology-mc2} but projected along z-direction.}
\label{fig-appendix-proj-mc2z}
\end{figure}

\end{document}